\documentclass[journal]{IEEEtranTCOM}

\normalsize

\ifCLASSINFOpdf
\else
\fi
\usepackage{subcaption} 
\usepackage{comment}
\usepackage{systeme}
\newcommand{\ycl}[1]
{\textcolor{purple}{#1}}

\usepackage{physics}
\usepackage{balance}
\newcommand{\tsp}
{\mathsf{\scriptstyle{T}}}
\newcommand{\bs}[1]{\boldsymbol{#1}}
\newcommand{\diff}{\ \mathrm{d}}

\newcommand*{\mc}{\mathcal}

\newcommand{\R}{\mathbb{R}}

\newcommand{\E}{\mathbb{E}}
\usepackage{amsthm} 

\theoremstyle{definition}
\newtheorem*{lma}{Lemma}

\theoremstyle{definition}
\newtheorem{thm}{Theorem}

\theoremstyle{definition}

\theoremstyle{definition}
\newtheorem{cor}{Corollary}

\usepackage{array}
\usepackage{makecell}

\usepackage{blindtext}
\usepackage{xcolor}

\usepackage{cite}
\usepackage{amsmath,amssymb,amsfonts}
\usepackage{algorithmic}
\usepackage{graphicx}
\usepackage{textcomp}
\usepackage{xcolor,soul}
\def\BibTeX{{\rm B\kern-.05em{\sc i\kern-.025em b}\kern-.08em
    T\kern-.1667em\lower.7ex\hbox{E}\kern-.125emX}}

\usepackage{mathrsfs}
\usepackage{amsthm,cite,url,graphicx,booktabs,lipsum,color,bm,caption,subcaption,soul}

\begin{document}
\title{Characterizing First Arrival Position Channels: Noise Distribution and Capacity Analysis}
\author{
Yen-Chi~Lee$^*$,~
Yun-Feng~Lo,~
Jen-Ming~Wu,~\IEEEmembership{Member,~IEEE,}
        and~Min-Hsiu~Hsieh,~\IEEEmembership{Senior~Member,~IEEE}


\thanks{
This manuscript refines and extends the concepts from our unpublished preprints \cite{lee2022arrival,lo2023vdfap}, presenting a more comprehensive and integrated narrative of the research.
}


\thanks{Yen-Chi Lee, Jen-Ming Wu and Min-Hsiu Hsieh are all affiliated with Hon Hai (Foxconn) Research Institute, Taipei, Taiwan. Emails: yenchilee1925@gmail.com, jen-ming.wu@foxconn.com, min-hsiu.hsieh@foxconn.com}
\thanks{Yun-Feng Lo is affiliated with Georgia Institute of Technology, Atlanta, Georgia, United States. Email: ylo49@gatech.edu}
}
\maketitle
\begin{abstract}
\color{black}
This paper introduces a novel mathematical model for Molecular Communication (MC) systems, utilizing First Arrival Position (FAP) as a fundamental mode of information transmission. We address two critical challenges: the characterization of FAP density and the establishment of capacity bounds for channels with vertically-drifted FAP. Our method relate macroscopic Partial Differential Equation (PDE) models to microscopic Stochastic Differential Equation (SDE) models, resulting in a precise expression that links FAP density with elliptic-type Green's function. This formula is distinguished by its wide applicability across any spatial dimensions, any drift directions, and various receiver geometries. We demonstrate the practicality of our model through case studies: 2D and 3D planar receivers. The accuracy of our formula is also validated by particle-based simulations. 
Advancing further, the explicit FAP density forms enable us to establish closed-form upper and lower bounds for the capacity of vertically-drifted FAP channels under a second-moment constraint, significantly advancing the understanding of FAP channels in MC systems. 



\color{black}


\end{abstract}

%
\IEEEpeerreviewmaketitle
\section{Introduction}
 \label{section:intro}


\color{black}

The challenge of transmitting information over a distance while maintaining the fidelity of the information has been present throughout human history, from ancient times to the present day. While modern communication systems have addressed this problem using electromagnetic signals, such techniques encounter limitations in nano-scale applications due to restrictions on wavelength, antenna size, and energy \cite{nakano2013molecular, guo2015molecular}. To circumvent these restrictions, molecular signals have been proposed as a viable alternative, particularly for nanonetwork applications \cite{akyildiz2008nanonetworks,pierobon2010physical}. 
This has given birth to a new field of communication called molecular communication (MC).


\subsection{Background on Molecular Communication}
In MC systems, information is conveyed through tiny molecules, referred to as message molecules (MMs), that act as information carriers \cite{farsad2016comprehensive}. To transport these MMs through the physical channel toward a receiver, a propagation mechanism is necessary. Various forms of propagation mechanisms exist, such as diffusion-based
\cite{pierobon2012capacity},
flow-based \cite{kadloor2012molecular}, or an engineered transport system like molecular motors \cite{moore2006design}.
Among these mechanisms, diffusion-based propagation
combined with advection (or somtimes chemical reaction 
\cite{bi2020chemical}), 
has been the prevalent approach considered in the MC literature
thus far. In this study, we will focus on \emph{diffusion-based} (also known as \emph{diffusive}) MC systems.
Note that we use the terms ``molecules'' and ``particles'' interchangeably to refer to the MMs, as their shape is not relevant to our discussion.

In diffusive MC systems, information can be transmitted by modulating various physical properties of the MMs \cite{kuran2020survey}. 
Once these signaling molecules reach the vicinity of the receiver, they can be detected and processed to extract the required information for detection and decoding \cite{jamali2019channel}. 
The reception mechanism of an MC receiver can be 
classified into two categories: i) \text{passive reception}, and ii) \text{active reception}
\cite{noel2016active}.
A widely used simple active reception model is the fully-absorbing receiver \cite{yilmaz2014three}.
\color{black}
For our following discussion, we will focus on the fully-absorbing receiver, which often referred to as an ``absorbing receiver'' for brevity. Note that this assumption is also commonly employed in recent research on molecular MIMO \cite{koo2016molecular} and molecular index modulation \cite{gursoy2019pulse,gursoy2020molecular}.
\color{black}


\color{black}
Absorbing receivers harness the information extracted during the initial contact between MMs and the receiver boundary.
The stochastic properties of MMs can be captured by first-hitting process (see \cite{koo2016molecular}).
The modulation schemes \cite{kuran2020survey} applicable to absorbing receivers can be roughly categorized into three types: timing-based \cite{srinivas2012molecular}, position-based \cite{lee2016distribution,pandey2018molecular}, and joint timing-position-based schemes \cite{akdeniz2018molecular}.
\color{black}
In this study, we delve deep into position-based information type  and explore a novel channel family known as First Arrival Position (FAP) channels.
\color{black}



\subsection{Main Results of this Work}
\color{black}
In this paper, we extend upon the materials introduced in our previously unpublished preprints \cite{lee2022arrival,lo2023vdfap}. 
We have dedicated substantial effort to refine and integrate the fundamental ideas and discoveries from these earlier versions. 
In addition, we have enriched the content to bolster the motivation behind this study and have proposed potential applications (Section~I-B).
Furthermore, in this journal version, we have performed particle-based simulations to verify the validity of our derived FAP density formula, and performed graphical illustrations of our derived capacity bounds with respect to key parameters.
Last but not least,
we have organized our key results about capacity in a theorem-style format (see Section~V) to facilitate easy comprehension for readers. 

\color{black}
The main contributions of this paper are as follows:
\begin{itemize}
\item \textbf{Modeling FAP-type MC systems:} By imposing some ideal assumptions, we model FAP-type MC systems as additive vector channels (see Eq.~(4)), and the particle's evolution is captured by an It\^{o} diffusion. Please see Fig.~1 and Fig.~2.
\item \textbf{FAP Channel Characterization:} 
We introduce an unified framework (Section~III) capable of calculating the FAP density in diffusive MC systems. This framework is applicable to MC systems of arbitrary spatial dimension, arbitrary drift direction, and various receiver shapes including the commonly used sphere and plane shapes.
\item \textbf{Upper and Lower Bounds on Capacity:} 
Under a second-moment constraint, we derive the upper and lower bounds for VDFAP channel capacity. To the best of our knowledge, this is the first work in MC to analytically explore the fundamental data transmission rate (i.e., channel capacity) for FAP-type MC channels.
\end{itemize}
\subsection{Organization of this Paper}
The remainder of this paper is structured as follows.
In Section~\ref{sec:sys}, we present the physical model and system assumptions used in our analysis. 
Section~\ref{section:generator} introduces our novel unified approach for finding FAP density.
In Section~\ref{sec:cf}, we examine the characteristic function of the VDFAP noise distribution, as well as its weak stability property. We then use the derived formulas for moments and the weak stability property to establish lower and upper bounds for the capacity of VDFAP channels in Section~\ref{sec:bounds}.
\textcolor{black}{
Numerical results are provided in Section~VI.}
Finally, we summarize our results and provide concluding remarks in Section~\ref{sec:conc}.

\section{System Model}\label{sec:sys}

\subsection{Physical Model}
There are two different perspectives for modeling diffusion phenomenon in MC systems. The macroscopic viewpoint utilizes the diffusion equation, also known as the heat equation (both are partial differential equation (PDE) models), to depict the evolution of the concentration $c(\mathbf{x},t)$ of message molecules. 
Based on Fick's law of diffusion 
\cite{jamali2019channel},
the temporal evolution of $c(\mathbf{x},t)$, assuming no chemical reactions, can be described as \cite[eq.~(17)]{jamali2019channel}:
\begin{equation}
\partial_{t} c\left(\mathbf{x},t;\mathbf{x}_{0}\right)
+
\nabla \cdot \left(
\mathbf{v}(\mathbf{x}, t)c\left(\mathbf{x},t;\mathbf{x}_{0}\right)
\right)
=D \Delta c\left(\mathbf{x},t;\mathbf{x}_{0}\right),
\label{ad-eq}
\end{equation}
where $\mathbf{x}_{0}$ denotes the emission point, $\mathbf{v}$ denotes the background velocity field, 
$\nabla \cdot$ represents the divergence operator, and $\Delta\equiv \nabla^2$ is the Laplacian operator\textcolor{black}{\cite[Section~12.11]{kreyszig2011advanced}}. Here, $D$ is the diffusion coefficient \cite{jamali2019channel}, which we assume to be constant with respect to space and time.
The value of $D$ is dependent on factors such as temperature, fluid viscosity, and the molecule's Stokes radius, as outlined in \cite{farsad2016comprehensive}.
Note that the shorthand notations $\partial_t$ stands for $\frac{\partial}{\partial t}$ and $\partial_{x_j}$ stands for $\frac{\partial}{\partial {x_j}}$.




In contrast, from a microscopic perspective, a suitable model for an 
individual trajectory $\mathbf{X}_t$ of a message molecule is an It\^{o} diffusion process\textcolor{black}{\cite[Section~7.1]{oksendal2013stochastic}}. 
We will assume that the parameters in the diffusion channel vary slowly over time, allowing the MC channel to be viewed as an approximately time-invariant system \cite{jamali2019channel}. This assumption corresponds to the time-homogeneous It\^{o} diffusion model.
Namely, a \emph{time-homogeneous It\^{o} diffusion} process is a stochastic process that satisfies a stochastic differential equation (SDE) of the
form \cite[Section~6.2.1]{calin2015informal}:
\begin{equation}
\diff \mathbf{X}_t = \mathbf{b}(\mathbf{X}_t)\diff t + \sigma(\mathbf{X}_t) \diff \mathbf{B}_t,
\label{eq:IP}
\end{equation}
where $\mathbf{B}_t$ is a $\text{D}$-dimensional standard
Brownian motion. The first term, $\mathbf{b}(\mathbf{X}_t)\ \mathrm{d}t$, is the deterministic component, which describes the drift effect induced by an external potential field.
The second term, $\sigma(\mathbf{X}_t)\  \mathrm{d}\mathbf{B}_t$, represents the random fluctuations caused by the constant bombardment of background molecules.
Notice that $\mathbf{b}(\cdot)$ is a $\text{D}$-dimensional vector, and $\sigma(\cdot)$ is a $\text{D}\times \text{D}$ matrix. Both are functions of the spatial variable $\mathbf{x}$.

These two models, PDE and SDE, possess an important relation.
The microscopic perspective via the SDE \eqref{eq:IP} can be connected to the macroscopic perspective via the PDE \eqref{ad-eq} through the Fokker-Planck equation, also known as the Kolmogorov forward equation\textcolor{black}{\cite[Chapter~8]{oksendal2013stochastic}}. Specifically, for an It\^{o} diffusion process $\mathbf{X}_t$ satisfying \eqref{eq:IP} for $t>0$ and the initial condition $\mathbf{X}_0=\mathbf{x}_0$, the probability density of $\mathbf{X}_t$, denoted by $p(\mathbf{x},t;\mathbf{x}_0)$, satisfies the Kolmogorov forward equation:
\begin{align}
\begin{split}
    \partial_t p(\mathbf{x},t;\mathbf{x}_0)
    &=
    - \sum_{j=1}^{\text{D}} \partial_{x_j}
    \left[ \mathbf{b}(\mathbf{x}) p(\mathbf{x},t;\mathbf{x}_0) \right]
    \\
    &\quad
    + \sum_{j=1}^{\text{D}} \sum_{k=1}^{\text{D}}
    \partial_{x_j} \partial_{x_k}
     \left[ D_{jk}(\mathbf{x}) p(\mathbf{x},t;\mathbf{x}_0) \right]
     \label{eq:Fokker-Planck}
\end{split}
\end{align}
where $b_j(\mathbf{x})$ denotes the $j$-th component of $\mathbf{b}(\mathbf{x})$ and $D_{jk}(\mathbf{x})$ is the $(j,k)$-th entry of the matrix $D(\mathbf{x}):=\tfrac{1}{2}\sigma(\mathbf{x})\sigma(\mathbf{x})^\tsp$. To establish a connection between \eqref{eq:Fokker-Planck} and \eqref{ad-eq}, we make two assumptions that are common in the MC literature \cite{farsad2016comprehensive,jamali2019channel}:
    i) the background velocity field is constant with respect to time, i.e., assuming $\mathbf{v}(\mathbf{x},t)=\mathbf{v}(\mathbf{x})$ in the macroscopic PDE model;
    ii) the diffusion coefficient $D$ is constant throughout space and time. That is, we assume the matrix $\sigma(\mathbf{x})$ to be a scalar multiple of the identity matrix, where the scalar is a constant. To simplify notations, we also denote this scalar constant by $\sigma$. 
Under these two assumptions, we can establish equivalence between \eqref{ad-eq} and \eqref{eq:Fokker-Planck} by identifying the following quantities: $c=p$, $\mathbf{v}(\cdot)=\mathbf{b}(\cdot)$ and $D=\sigma^2/2$.\footnote{Note that since \eqref{ad-eq} and \eqref{eq:Fokker-Planck} are \emph{linear} PDEs, any scalar multiple of a solution is still a solution. In concentration-encoded MC where MMs move independently, the relation $c=N_\text{total}\cdot p$ holds ($N_\text{total}$ denotes the total number of MMs released) \cite{jamali2019channel}.}

\color{black}
\subsection{Communication Model} \label{subsec:comm-model}

\begin{figure}[!t]  
\centerline{\includegraphics[width=0.5\textwidth]
{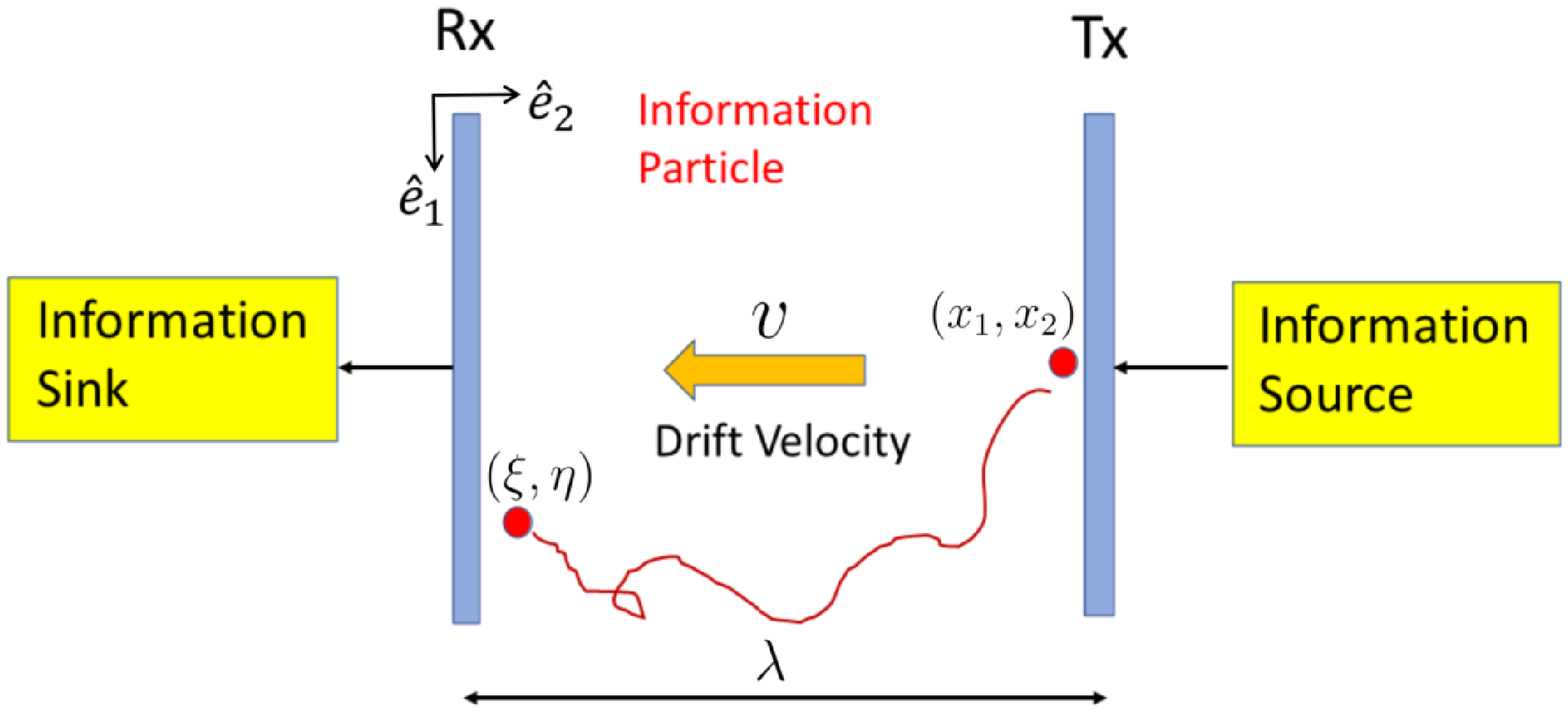}}
\caption{
This figure illustrates a 2D FAP channel with line-shaped Tx and Rx, where the Tx is assumed to be transparent, allowing particles to move through it without experiencing any force. The emission point is located on the Tx line.
\textcolor{black}{Note that the two red particles are highlighted as exemplary particles to indicate the emission position $(x_1, x_2)$ and the hitting position $(\xi, \eta)$, respectively.}
}
\label{fig:2Da}
\end{figure}
\begin{figure}[!t]
\centerline{\includegraphics[width=0.5\textwidth]
{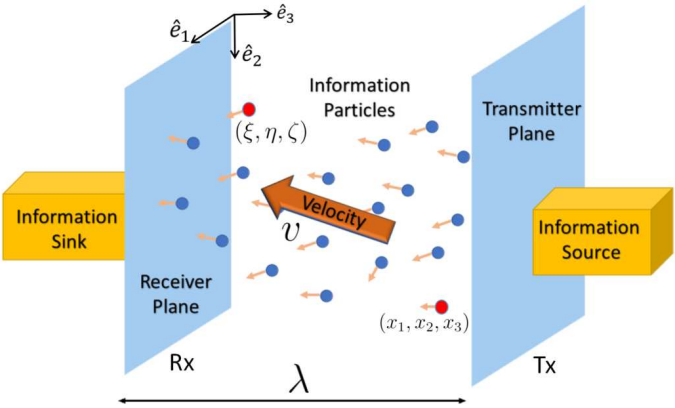}}
\caption{
This figure illustrates a 3D FAP channel with plane-shaped Tx and Rx, where the Tx is assumed to be transparent, allowing particles to move through it without experiencing any force. The emission point is located on the Tx plane. 
\textcolor{black}{Note that the two red particles are highlighted as exemplary particles to indicate the current position $(x_1, x_2, x_3)$ and the hitting position $(\xi, \eta, \zeta)$, respectively.}
}
\label{fig:3Da}
\end{figure}
In this paper, we investigate a diffusive MC system, which employs the emission positions of MMs in a $\text{D}$-dimensional fluid medium to convey information. In practice, $\text{D}=2,3$ are commonly selected. See Fig.~\ref{fig:2Da} and Fig.~\ref{fig:3Da}.
We consider a scenario where there is a constant velocity field \cite{kadloor2012molecular} in the ambient space. This corresponds to further assuming $\mathbf{v}(\cdot)=\mathbf{b}(\cdot)=\mathbf{v}$, where $\mathbf{v}$ is a constant vector. We call $\mathbf{v}$ the drift (vector) hereafter.
Additionally, we assume all the MMs are of the same type \cite{farsad2015stable}.
The diffusion effects of MMs are captured by a diffusion coefficient $D=\sigma^2/2$, 
which is assumed to be a constant throughout space and time \cite{farsad2016comprehensive}. We call this the isotropic assumption on the fluid medium. 

The MC system we considered comprises a transmitter (Tx) and a receiver (Rx), which are modeled as parallel $(\text{D}-1)$-dimensional hyperplanes (hereafter referred to as \emph{planes}) separated by a distance $\lambda>0$. Without loss of generality, we can set the Tx plane at coordinate $x_{\text{D}}=\lambda$ and the Rx plane at coordinate $x_{\text{D}}=0$. 


\color{black}
In this study, we assume some simplified assumptions to our FAP MC system.
\begin{itemize}
\item[(A1)] The movement of every MM is independent. (Most MC works adopt this assumption.)
\item[(A2)] Upon first arrival at the receiver plane, MMs are captured and removed from the system. (Most MC works using an fully-absorbing receiver adopt this assumption.)
\item[(A3)] The receiver can perfectly measure the first arrival positions of the MMs. (Note: We add this assumption in order to have a tractable analysis for channel capacity. However, the main results in Section~III and Section~IV do not depend on this assumption.)
\end{itemize}
\color{black}

Although the 
ambient space is D-dimensional, the position information is $d$-dimensional, where $d:=\text{D}-1$ for brevity and consistency throughout this paper. The first arrival position $\bs{Y}$ of a molecule released at position $\bs{X}$ can be expressed as
\begin{equation}\label{fap-channel}
\bs{Y}=\bs{X}+\bs{N},
\end{equation}
where $\bs{N}$ denotes the deviation (i.e., noise) of the first arrival position. 
The probability density function of $\bs{N}$ is defined to be the \emph{FAP density} \cite{lee2016distribution,pandey2018molecular},
and the additive channel \eqref{fap-channel} is defined to be the \emph{FAP channel}.

\color{black}

Please note that the realization of the random vector \(\bs{N}\) is represented by the lowercase letter \(\bs{n}\). In instances where \(\bs{n}\) resides in a \(\R^d\) space, we express it as $\bs{n}=(n_1, n_2, \ldots, n_d)$, which can also be equivalently denoted as $[n_1, n_2, \ldots, n_d]$.
The same convention will also be applied to other random vectors, such as $\bs{X}$ and $\bs{Y}$.

\color{black}

\subsection{The FAP Noise Distribution}\label{subsec:fap}
\color{black}
In \cite{pandey2018molecular,lee2016distribution}, the 2D and 3D FAP densities were separately derived based on different methods.
By the framework proposed in this paper, we can unify the derivations in different scenarios into a standard procedure. Our new derivation, which recovers the 
FAP formulas in 
\cite{pandey2018molecular,lee2016distribution}, is presented 
in Appendix~\ref{sec:case}.
Notice that the special function $K_{\frac{3}{2}}$ has an elementary form\textcolor{black}{\cite[Section 8.432]{gradshteyn2014table}}:
\begin{equation}\label{k32}
    K_{\frac{3}{2}}(x)
    =
    \sqrt{\dfrac{\pi}{2}}e^{-x}\left(
    \dfrac{1+x}{x^{\frac{3}{2}}}
    \right),
\end{equation}
where $K_{\nu}(\cdot)$ denotes the order-$\nu$ modified Bessel function of the second kind \cite{zill2020advanced}.
By denoting $\bs{n}=\bs{y}-\bs{x}$, we can merge the two FAP densities, for $\text{D}=2$ 
and $\text{D}=3$,  
into a single expression:
\begin{align} \label{eq:dD-FAP}
\begin{split}
    f^{(d)}_{\bs{N}}(\boldsymbol{n}; \mathbf{u},\lambda)
    =\ 
    &2\lambda 
    \frac{ \| \mathbf{u} \|^{\tfrac{d+1}2}
    }{(2\pi)^{\tfrac{d+1}{2}}}
    \cdot e^{\boldsymbol{u}_{\text{par}}^\intercal \boldsymbol{n}-u_\text{D}\lambda} \\ 
    \cdot &\frac{K_{\tfrac{d+1}2}\left(\| \mathbf{u} \| \sqrt{\norm{\bs{n}}^2+\lambda^2}\right)}{\left(\sqrt{\norm{\bs{n}}^2+\lambda^2}\right)^{\tfrac{d+1}2}}
    ,
\text{\ for }
\bs{n} \in \mathbb{R}^d
\end{split}
\end{align}
where $\|\cdot\|$ denotes the Euclidean norm. \color{black}
We present here some special notational conventions used throughout this paper.
We write the $\text{D}$-dimensional drift vector $\mathbf{v}=[v_1,\ldots,v_\text{D}]^\tsp$ as $\mathbf{v}=[\boldsymbol{v}_\text{par}^\tsp,v_\text{D}]^\tsp$, where $\boldsymbol{v}_\text{par}=[v_1,\ldots,v_{d}]^\tsp$ contains the drift components parallel to the Tx and Rx planes, and $v_{\text{D}}$ is the drift component perpendicular to the Tx and Rx planes. As a rule of thumb, we use \textbf{bold fonts} to stand for (column) vectors; slanted \textbf{\textit{vectors}} are $d$-dimensional, while non-slanted \textbf{vectors} are $\text{D}$-dimensional. The operator $(\cdot)^\tsp$ represents transposition.
We also introduce the notation
$
\mathbf{u}
=
\frac{\mathbf{v}}{\sigma^2}
$
, $\boldsymbol{u}_\text{par}=
\frac{\boldsymbol{v}_\text{par}}{\sigma^2}$ and $u_j=\frac{v_j}{\sigma^2}$ for each $j\in\{1,\ldots,\text{D}\}$. 
These $\{u_j\}_{j=1}^\text{D}$ can be interpreted as \emph{normalized drift}: they are the ratio of the drift ``strength" $v_j$ over the diffusion ``strength" $\sigma^2$. Since $v_j$ has the unit of $\mu \mathrm{m}/s$ and $\sigma^2=D/2$ has the unit of $\mu \mathrm{m}^2/s$, the normalized drift $u_j$ has the unit of $\mu \mathrm{m}^{-1}$.
\color{black}
 

For the channel capacity investigation, we focus on a sub-family of FAP (noise) distributions with two specific properties:
\begin{itemize}
    \item[i)] The parallel drift components are zero, i.e., $\boldsymbol{v}_\text{par}=\boldsymbol{0}$.
    \item[ii)] The \emph{vertical drift} (VD) component points from the Tx to the Rx, i.e., $v_\text{D}<0$. (Intuitively, this vertical drift helps the transmission of information.) 
\end{itemize} 
We name this sub-family as vertically-drifted first arrival position (VDFAP) distributions because only the vertical drift component is non-zero. We can set $\boldsymbol{u}_\text{par}=\boldsymbol{0}$ so that $\|\mathbf{u}\|$ in \eqref{eq:dD-FAP} becomes $|u_\text{D}|=-u_\text{D}$.
Abbreviating $u_\text{D}$ as $u$, the VDFAP densities can be expressed as:
\begin{align} \label{eq:dD-VDFAP}
\begin{split}
f^{(d)}_{\bs{N}}(\boldsymbol{n}; u,\lambda)
=\ 
&2\lambda \left(\frac{\vert u \vert}{\sqrt{2\pi}}\right)^{d+1} 
\cdot
e^{\lambda\vert u \vert}\\
\cdot
&\frac{K_{\tfrac{d+1}2}\left(\vert u \vert\sqrt{\norm{\bs{n}}^2+\lambda^2}\right)}{\left(\vert u \vert\sqrt{\norm{\bs{n}}^2+\lambda^2}\right)^{\tfrac{d+1}2}}
,
\text{\ for }
\bs{n} \in \mathbb{R}^d.
\end{split}
\end{align}
We also denote the VDFAP densities by $f^{(d)}_{\bs{N}}(\boldsymbol{n})$ when the parameters $u,\lambda$ are clear from the context.
As a shorthand, we denote $\bs{N}\sim\text{VDFAP}^{(d)}(u,\lambda)$ if a $d$-dimensional random vector $\bs{N}$ follows the VDFAP distribution defined by \eqref{eq:dD-VDFAP} with parameters $u<0$ and $\lambda>0$.
Note that we refer to a random variable $X\in\mathbb{R}$ as a ``one-dimensional random vector" to maintain consistency in terminology.
\color{black}
\section{A Novel Framework for Efficiently Determining FAP Density} \label{section:generator}

In this section, we propose a novel and unified framework for determining the FAP density. To achieve this goal, we introduce several stochastic analytical tools, such as the 
generator of diffusion semigroup 
and Dynkin's formula\textcolor{black}{\cite[Section~9.3]{calin2015informal}}.

\color{black}
Our new method for calculating the FAP density can be divided into three steps as summarized in Table~\ref{table:compare-gpt}.
Specifically, the FAP density can be computed as $f_{\mathbf{Y}|\mathbf{X}}(\mathbf{y}|\mathbf{x})=
\big| \frac{\partial G(\mathbf{x},\mathbf{y})}{\partial \mathbf{n}_{\mathbf{y}}} \big|$ once the elliptic Green's function $G(\mathbf{x},\mathbf{y})$ is known.
Note that the ``previous method" in Table~\ref{table:compare-gpt} refers to those methods that tackle the diffusion equation \eqref{ad-eq} directly, such as \cite{pandey2018molecular}.

For the previous method, a parabolic-type PDE (which is time-dependent) is used, while the new method removes the time-dependency by considering the generator of the diffusion semigroup. 
As a consequence, using the previous method often requires doing integration with respect to time, which is unnecessary while applying our new method.

\begin{table*}[t]
\caption{Comparison Between Previous and New Method of Calculating FAP Density}
\centering
\begin{tabular}{|c|p{6.75cm}|p{6.75cm}|}
\hline
& \textbf{Previous Method}
& \textbf{New Method \textcolor{black}{(we propose in this work)}} \\ \hline
\textbf{Step 1} & Calculates the free space Green's function for parabolic-type\textcolor{black}{\cite[Chapter~3]{polyanin2001handbook}} PDE. & Removes time-dependency via considering the generator of the diffusion semigroup. \\ \hline
\textbf{Step 2} & Accommodates the free space solution to absorbing boundary conditions (e.g., using the image method). & Solves elliptic-type\textcolor{black}{\cite[Chapter~8]{polyanin2001handbook}} boundary value problem with appropriate boundary conditions. \\ \hline
\textbf{Step 3} & Calculates the flux density (see \cite{pandey2018molecular}) and does integration with respect to time. & Obtains FAP density directly from the Green's function $G(\mathbf{x},\mathbf{y} )$. No time-integration is required. \\ \hline
\end{tabular}
\label{table:compare-gpt}
\end{table*}

In the remainder of this section, we will establish the validity of the proposed FAP density computing framework. 


\subsection{Infinitestimal Generator of It\^{o} Diffusion}

\color{black}

Recall that a $\text{D}$-dimensional It\^{o} diffusion
is a stochastic process that satisfies \eqref{eq:IP}:
$
\diff \mathbf{X}_t = \mathbf{b}(\mathbf{X}_t)\diff t + \sigma(\mathbf{X}_t) \diff \mathbf{B}_t
.
$
In this subsection, we will first assume that $\mathbf{b}(\cdot)$ and $\sigma(\cdot)$ are constant only with respect to time, but not to space, allowing the corresponding channel to be regarded as a time-invariant system. 
We will make further assumptions on $\sigma(\cdot)$ as the deductions progress. Nevertheless, for the purpose to calculate FAP density in specific scenarios 
(as presented in Appendix~\ref{sec:case}),
we only need the stronger assumptions as stated in Section~\ref{sec:sys}.


\color{black}
For each It\^{o} diffusion $\mathbf{X}_t$, a corresponding operator called the infinitesimal generator, or simply generator, can be associated. 
Let $\mathcal{D}(A)$ denote the domain of the generator $A$ \cite{stone1932one}. We define the notation $\E^{\mathbf{x}}[\cdot]$ as taking expectation conditioned on $\mathbf{X}_{0}=\mathbf{x}$, so that
\color{black}
\begin{equation}\label{eq:expect-defi}
\mathbb{E}^{\mathbf{x}}[f(\mathbf{X}_t)] := \mathbb{E}[f(\mathbf{X}_t)|\mathbf{X}_0=\mathbf{x}].
\end{equation}
\color{black}
The generator $A$ (which operates on function $f$) of a process $\mathbf{X}_t$ can be defined as
\begin{equation}
Af=A\{f(\mathbf{x})\}:=\lim\limits_{t\searrow 0} \dfrac{\mathbb{E}^{\mathbf{x}}[f(\mathbf{X}_t)]-f(\mathbf{x})}{t}
\text{\ for any\ } f\in \mathcal{D}(A).
\label{eq:def-generator}
\end{equation}
For time-homogeneous It\^{o} processes $\mathbf{X}_t$, the time evolution is a Markov process.
By letting
$
T_t f := \E^{\mathbf{x}}[f(\mathbf{X}_t)],
$
we can obtain a semigroup of operators $\mathcal{T}=\{T_t\}_{t\geq 0}$.
Notice that the term 
\begin{equation}
Af = \lim\limits_{t\searrow 0} \dfrac{T_t f - f}{t}
\end{equation}
can be interpreted as the rate of linear increment with respect to $t$ through the semigroup evolution.
\color{black}

\color{black}
To explicitly calculate the generator $A$ based on \eqref{eq:IP},
we can assume that $f$ is of class $C^2$, which means that it has both a continuous first derivative and a continuous second derivative. 
Using Taylor expansion \cite[Formula (9.2.2)]{calin2015informal}
and It\^{o}'s formula\textcolor{black}{\cite[Chapter~4]{oksendal2013stochastic}},
we have:
\begin{align}
\begin{split}
\diff f\left(\mathbf{X}_t\right)=\ & (\nabla f)^\tsp(\mathrm{d}\mathbf{X}_t)  +\frac{1}{2} (\mathrm{d} \mathbf{X}_t)^\tsp
\text{Hess}(f)
(\mathrm{d}\mathbf{X}_t)
\\
=\ &\left[(\nabla f)^\tsp (\mathbf{b}\left(\mathbf{X}_t\right))+\tr(\text{Hess}(f)\frac{\sigma\sigma^\tsp \left(\mathbf{X}_t\right)}{2}
)\right] \mathrm{d} t\\
&+ (\nabla f)^{\tsp} \sigma\left(\mathbf{X}_t\right) \mathrm{d} \mathbf{B}_{t},
\end{split}
\label{eq:te_modified}
\end{align}
where $\tr(\cdot)$ denotes the trace operation, $\text{Hess}(f)$ denotes the Hessian of $f$ and the matrix notation $\sigma\sigma^\tsp \left(\mathbf{X}_t\right)$ is short for $\sigma\left(\mathbf{X}_t\right) \left(
\sigma \left(\mathbf{X}_t\right)
\right)^\tsp$. Note that in \eqref{eq:te_modified}, all the gradient and Hessian terms are evaluated at $\bs{X}_t$. After taking expectation to both sides of \eqref{eq:te_modified}, and then plugging into \eqref{eq:def-generator}, we use the fact that $\mathbb{E}[\mathrm{d} \mathbf{B}_t]=0$ to obtain an explicit expression for the generator $A$ as:
\begin{equation}
A\{f(\mathbf{x})\}=
\mathbf{b}(\mathbf{x})^\tsp
(\nabla f(\mathbf{x}))
+\dfrac{\sigma^2(\mathbf{x})}{2}\Delta f(\mathbf{x}),
\label{eq:ItoGe_modified}
\end{equation}
where $\mathbf{x}$ can be viewed as the starting point of a diffusion process. From \eqref{eq:ItoGe_modified}, we can see that $A$ is a second-order differential operator of elliptic-type. 
Notice that in \eqref{eq:ItoGe_modified} we have assumed that the matrix $\sigma(\cdot)$ is a scalar multiple of the identity matrix. To simplify the notation, we also denoted this scalar function by $\sigma(\cdot)$ in \eqref{eq:ItoGe_modified}.
\color{black}

We proceed to consider a boundary value problem (BVP) with an unknown function $\phi\in C^2$, given by:
\begin{equation}\label{eq:BVP1-1}
\left\{\begin{array}{ll}
A\phi=0 & \text { in } \Omega \\
\phi=g & \text { on } \partial \Omega
\end{array}\right.
,
\end{equation}
where $\Omega$ is the domain in which the MMs can diffuse, and $\partial \Omega$ is the boundary of the domain.
In \eqref{eq:BVP1-1}, $A$ denotes a partial differential operator and $g$ is the prescribed boundary data, which is also a $C^2$ function. Note that for elliptic-type partial differential operators, representation formulas for commonly encountered boundary conditions can be directly found in PDE handbooks such as \cite{polyanin2001handbook}. 


\subsection{A Novel Relation Between FAP Density and Elliptic Green's Function}\label{subsection:D}

\color{black}
Recall the infinitesimal generator defined in \eqref{eq:def-generator}. By rearranging the terms in \eqref{eq:def-generator}, we can obtain the following integral expression:
\begin{equation}\label{generator-int}
\E^{\mathbf{x}}\left[f\left(\mathbf{X}_t\right)\right]=f(\mathbf{x})+\E^{\mathbf{x}}\left[\int_{0}^{t} A \{f\left(\mathbf{X}_s\right)\} \diff s\right]
.\end{equation}
It is important to note that the variable $t$ appearing in \eqref{generator-int} is deterministic, not a random variable.
\color{black}


\color{black}
In the field of stochastic analysis, 
Dynkin's formula relates the expected value of a function $f$ of a Markov process at a stopping time $\tau$ \cite[Definition~8.2.1]{shreve2004stochastic}
to the initial value $f(\mathbf{x})$ of the function and an integral involving the generator $A$ of the process.
Technically, we have to make some assumptions on $f$.
As we are considering a second-order differential operator \eqref{eq:ItoGe_modified}, it is common to require $f$ to be twice continuously differentiable (i.e., $f\in C^2$).
Letting $\tau$ be a stopping time such that $\E^{\mathbf{x}}[\tau]<+\infty$, the Dynkin's formula can be expressed as:
\begin{equation}
\E^{\mathbf{x}}\left[f\left(\mathbf{X}_{\tau}\right)\right]=f(\mathbf{x})+\E^{\mathbf{x}}\left[\int_{0}^{\tau} A \{f\left(\mathbf{X}_s\right)\} \diff s\right].
\label{formula-dynkin}
\end{equation}
Notably, $\tau$ in \eqref{formula-dynkin} is now a random variable. Additional details regarding this formula can be found in \cite[eq.~(9.3.10)]{calin2015informal}.
\color{black}

\color{black}
Let $g$ denote a smooth function defined on the boundary $\partial \Omega$. 
We can express the expected value of $g(\mathbf{X}_\tau)$ as follows:
\begin{align}
\begin{split}
\E^{\mathbf{x}}\big[g(\mathbf{X}_\tau)\big]
&=\E\Big[g(\mathbf{X}_\tau)\Big\vert \mathbf{X}_0=\mathbf{x}\Big]
=\int_{\partial \Omega} f_{\mathbf{Y}|\mathbf{X}}(\mathbf{y}| \mathbf{x})g(\mathbf{y})\diff S_\mathbf{y},
\label{eq:22}
\end{split}
\end{align}
where $\mathrm{d} S_\mathbf{y}$ is the magnitude of surface element (see\textcolor{black}{\cite[Section~10.6]{kreyszig2011advanced}}) at $\mathbf{y}$.
Here, $\mathbf{x}\in \Omega$ represents the starting point (corresponding to the emission point) of the diffusion, while the hitting position $\mathbf{y}$ belongs to the boundary, i.e., $\mathbf{y}\in\partial \Omega$.

The conditional probability density function $f_{\mathbf{Y}|\mathbf{X}}$ in \eqref{eq:22} can be interpreted as the distribution of the position of $\mathbf{X}_t$ at the stopping time $\tau$. In MC applications, $\mathbf{X}_t$
is the trajectory of a MM, and the stopping event corresponds to the capture by Rx. Thus, $f_{\mathbf{Y}|\mathbf{X}}$ is the desired FAP density on the receiver boundary $\partial \Omega$.
Our remaining task is to express $f_{\mathbf{Y}|\mathbf{X}}$ using the Green's function.
\color{black}

\color{black}
Suppose we have a solution $\phi(\mathbf{x})$ for the BVP in \eqref{eq:BVP1-1} with prescribed boundary data $g$ such that $A\phi=0$ inside $\Omega$. By setting $f(\mathbf{x})=\phi(\mathbf{x})$ in \eqref{formula-dynkin}, we obtain
$
\int_0^\tau A\{\phi(\mathbf{X}_s)\} \diff s =0.
$
Here, $\tau$ represents the first hitting time. The physical interpretation of this condition is that $\mathbf{X}_s$ is located inside $\Omega$ prior to the hitting event, which leads to $A\{\phi(\mathbf{X}_s)\}=0$ for $0<s<\tau$. Furthermore, since $\phi(\mathbf{x})$ coincides with $g(\mathbf{x})$ on the boundary, we can express the left-hand side of \eqref{formula-dynkin} as $\E^{\mathbf{x}}[g(\mathbf{X}_\tau)]$. Combining these two facts yields:
\begin{equation}
\E^{\mathbf{x}}[g(\mathbf{X}_\tau)]=\phi(\mathbf{x})+0=
\int_{\partial \Omega} f_{\mathbf{Y}|\mathbf{X}}(\mathbf{y}| \mathbf{x})g(\mathbf{y})\diff S_\mathbf{y}.
\label{main-1}
\end{equation}
This holds true for any $\mathbf{x}\in \Omega$.
\color{black}


\color{black}
For a general elliptic-type BVP defined in a domain $\Omega$ with boundary $\partial \Omega$, we have the following formula\textcolor{black}{\cite[Section~7.2.1]{polyanin2001handbook}}:
\color{black}
\begin{equation}
\phi(\mathbf{x})=\int_{\Omega} \Xi(\mathbf{y}) H_1(\mathbf{x}, \mathbf{y}) \diff V_{\mathbf{y}}+\int_{\partial \Omega} g(\mathbf{y}) H_2(\mathbf{x}, \mathbf{y}) \diff S_{\mathbf{y}}.
\label{general-green}
\end{equation}
\color{black}
Here, $\Xi$ represents the source term, $\mathrm{d} V_{\mathbf{y}}$ is the volume element\textcolor{black}{\cite[Section~10.7]{kreyszig2011advanced}}, $\mathrm{d} S_{\mathbf{y}}$ is the magnitude of surface element\textcolor{black}{\cite[Section~10.6]{kreyszig2011advanced}}, and $H_1(\mathbf{x}, \mathbf{y})$, $H_2(\mathbf{x}, \mathbf{y})$ depend on the type of boundary conditions under consideration. For MC systems without molecule reproduction or annihilation inside the channel, we can set $\Xi(\mathbf{y})=0$, and hence $H_1(\mathbf{x}, \mathbf{y})$ is irrelevant. 
Notice that BVP \eqref{eq:BVP1-1} has the Dirichlet-type boundary condition \cite[Chapter~13]{zill2020advanced}. According to\textcolor{black}{\cite[Section~7.2.1]{polyanin2001handbook}}, we have
\begin{equation}\label{eq:H2}
H_2(\mathbf{x},\mathbf{y})=-\dfrac{\partial G(\mathbf{x},\mathbf{y})}{\partial \mathbf{n}_\mathbf{y}}
\end{equation}
in the solution formula \eqref{general-green},
where $G$ represents the Green's function of BVP \eqref{eq:BVP1-1}.
The notation $\frac{\partial}{\partial \mathbf{n}_\mathbf{y}}$
means the directional derivative along the outward normal at the point $\mathbf{y}$ on $\partial \Omega$.
Letting $\Xi=0$ and plugging \eqref{eq:H2} into \eqref{general-green}, we obtain a simpler form:
\begin{equation}
\phi(\mathbf{x})=\int_{\mathbf{y}\in\partial \Omega}
\left|\dfrac{\partial G(\mathbf{x},\mathbf{y})}{\partial \mathbf{n}_\mathbf{y}}
\right|
g(\mathbf{y}) \diff S_{\mathbf{y}}.
\label{general-green-2}
\end{equation}
Note that we have added an absolute value since the term $\big| \frac{\partial G(\mathbf{x},\mathbf{y})}{\partial \mathbf{n}_\mathbf{y}}\big|$ has the physical meaning of conditional probability density, so it must be non-negative.
Finally, by comparing \eqref{main-1} and \eqref{general-green-2}, we can establish a fundamental relation between the FAP density and the elliptic-type Green's function:
\begin{equation}
f_{\mathbf{Y}|\mathbf{X}}(\mathbf{y}|\mathbf{x}) = \left| \dfrac{\partial G(\mathbf{x},\mathbf{y})}{\partial \mathbf{n}_\mathbf{y}}\right| \Bigg\vert_{\partial \Omega}
,
\label{eq:main}
\end{equation}
where $\big\vert_{\partial \Omega}$ means $\mathbf{y}$ is evaluated
on $\partial \Omega$.
Notice that \eqref{eq:main} gives a concise relation between the FAP density function and the Green's function. 
How to explicitly calculate the FAP density using \eqref{eq:main}
in specific scenarios is demonstrated 
in Appendix~\ref{sec:case}.
\color{black}

\section{\textcolor{black}{Properties of VDFAP Distributions via Frequency Domain Analysis}
}
\label{sec:cf}

To bound the VDFAP channel capacity under a second-moment constraint in Sec.~\ref{sec:bounds}, we conduct a frequency-domain analysis of the VDFAP distribution via its \emph{characteristic function} (CF). From the CF we investigate the first and second moments of VDFAP noise for analyzing the output second moment of VDFAP channels. Also, we discover a property we call \emph{weak stability} exhibited by VDFAP distributions. This property turns out to be essential for proving a lower bound for the VDFAP channel capacity.

Based on probability theory \cite[Section~3.3]{durrett2019probability}, a 
distribution 
can be characterized by its 
CF, which is the Fourier transform of its PDF, i.e., a frequency-domain representation of the distribution. Interestingly, via the frequency-domain viewpoint, it is easier to compute the moments as well as analyze the (weak) stability properties for VDFAP distributions.

For a random vector $\bs{N}$ following $\text{VDFAP}^{(d)}(u,\lambda)$, we denote its CF as
\begin{align}
    \Phi^{(d)}_{\bs{N}}(\boldsymbol{\omega};u,\lambda)
    =
    \mathbb{E}[
    \exp(i\bs{\omega}^\tsp\bs{N})], \text{\ for\ } \boldsymbol{\omega}\in\mathbb{R}^{d}
    ,
\label{eq:CF-def}
\end{align}
where $\mathbb{E}$ is the expectation operator and $i$ the imaginary unit. We also write the CF simply as $\Phi^{(d)}_{\bs{N}}(\boldsymbol{\omega})$ 
when the parameters are clear from the context.
To our best knowledge, we have derived a novel closed-form expression for the CF \eqref{eq:CF-def} of VDFAP when $d=1,2$, \textcolor{black}{detailed in Appendix~\ref{sec:appen-CF}}, as:
\begin{align}
\Phi^{(d)}_{\bs{N}}(\boldsymbol{\omega};u,\lambda)
=
\exp
\left(
-\lambda
\left(
\sqrt{\|\boldsymbol{\omega}\|^2+|u|^2}
-|u|
\right)
\right)
.
\label{eq:VDFAP-CF-revised}
\end{align}
\color{black}
Note that letting $|u|\to 0$ in \eqref{eq:VDFAP-CF-revised} reduces to the $d$-dimensional Cauchy CF form:
\begin{align}
\lim_{|u|\to 0}
\Phi^{(d)}_{\bs{N}}(\boldsymbol{\omega};u,\lambda)
=
\exp
\left(
-\lambda
\|\boldsymbol{\omega}\|
\right),
\label{eq:Cauchy-CF-revised}
\end{align}
which is discussed more in \cite{lee2022capacity}.


\color{black}

\color{black}
Via the closed-form CF expression \eqref{eq:VDFAP-CF-revised} of a VDFAP distribution, we are able to obtain, relatively easily, formulas for its moments in Subsec.~\ref{subsec:mean_and_cov}, as well as study its (weak) stability properties in Subsec.~\ref{subsec:stable}.
\color{black}

\subsection{The Moments}
\label{subsec:mean_and_cov}
\color{olive}
\color{black}

In this subsection, we aim to find the first two moments of VDFAP distributions. In particular, we derive the first moment $\mathbb{E}[\bs{N}]$ and the second moment\footnote{Whenever $\bs{N}$ has component random variables $N_1,\ldots,N_d$, we call $\mathbb{E}\left[\|\bs{N}\|^2\right]=\sum_{i=1}^d \mathbb{E}\left[|N_i|^2\right]$ the ``second moment" of $\bs{N}$.} $\mathbb{E}\left[\|\bs{N}\|^2\right]$ for their further use in analyzing the VDFAP channel capacity under a second-moment constraint.

In principle, moments can be calculated by integrating the product of monomials with the PDF. However, due to the complicated PDF \eqref{eq:dD-VDFAP}, it is not clear how to find closed-form expressions for these integrals. 
Due to its similarity to the moment-generating function, the CF is useful for the calculation of moments.
In particular, from \eqref{eq:CF-def}, the first moment of $\bs{N}\sim\text{VDFAP}^{(d)}(u,\lambda)$ can be calculated as
\begin{align}
    \mathbb{E}[\bs{N}]
    =
    -i
    \nabla 
    \Phi^{(d)}_{\bs{N}}(\boldsymbol{\omega})
    \bigg\vert_{\boldsymbol{\omega}=\boldsymbol{0}}
    \label{eq:VDFAP-mean}
\end{align}
(where $\nabla$ is the gradient operator, and $\big\vert_{\boldsymbol{\omega}=\boldsymbol{0}}$ denotes evaluation at $\boldsymbol{\omega}=\boldsymbol{0}$),
and its correlation matrix\footnote{We are interested in the correlation matrix because its trace is exactly the second moment: 
$
\Tr{
\mathbb{E}\left[\bs{N}\bs{N}^\tsp\right]
}
=
\mathbb{E}\left[
\Tr{
\bs{N} \bs{N}^\tsp
}
\right]
=
\mathbb{E}\left[
\Tr{
\bs{N}^\tsp \bs{N} 
}
\right]
=
\mathbb{E}\left[
\Tr{
\|\bs{N}\|^2 
}
\right]
=
\mathbb{E}
\left[
\|\bs{N}\|^2 
\right]
$,
where we applied the cyclic property of trace.
}
$\mathbb{E}
    \left[
    \bs{N}\bs{N}^\tsp
    \right]$ 
as
\begin{align}
    \mathbb{E}
    \left[
    \bs{N}\bs{N}^\tsp
    \right]
    =
    -
    \text{Hess}\Big(
    \Phi^{(d)}_{\bs{N}}\Big)(\boldsymbol{\omega})
    \bigg\vert_{\boldsymbol{\omega}=\boldsymbol{0}}.
\label{eq:VDFAP-corr}
\end{align}

To this end, we calculate the gradient of CF \eqref{eq:VDFAP-CF-revised} as
\begin{align}
    \nabla 
    \Phi^{(d)}_{\bs{N}}(\boldsymbol{\omega})
    =
    \frac{-\lambda \Phi^{(d)}_{\bs{N}}(\boldsymbol{\omega})}{\sqrt{\|\boldsymbol{\omega}\|^2+|u|^2}} \boldsymbol{\omega}
\label{eq:VDFAP-CF-grad}
\end{align}
and the Hessian matrix of CF as
\begin{align}
\begin{split}           \text{Hess}\Big( 
\Phi^{(d)}_{\bs{N}}\Big)(\boldsymbol{\omega})
=~& 
    \frac{-\lambda \Phi^{(d)}_{\bs{N}}(\boldsymbol{\omega})}{\sqrt{\|\boldsymbol{\omega}\|^2+|u|^2}}
    \mathbb{I}_{d}\\
    &+
    \frac{
    \lambda \Phi^{(d)}_{\bs{N}}(\boldsymbol{\omega})
    \left(
    1
    +
    \lambda
    \sqrt{\|\boldsymbol{\omega}\|^2+|u|^2}
    \right)
    }{(\|\boldsymbol{\omega}\|^2+|u|^2)^{\frac32}}
    \boldsymbol{\omega} \boldsymbol{\omega}^\tsp
    ,
\label{eq:VDFAP-CF-hess}
\end{split}
\end{align}
where $\mathbb{I}_{d}$ denotes the $d$-by-$d$ identity matrix.
From \eqref{eq:VDFAP-mean} and \eqref{eq:VDFAP-CF-grad}, we can deduce that 
\begin{align}
    \mathbb{E}[\bs{N}]
    =
    \bs{0}
    .
    \label{eq:VDFAP-mean-0}
\end{align}
 Alternatively, due to the radial symmetry of $\bs{N}$, we can also deduce that its first moment is zero. 
Also, from \eqref{eq:VDFAP-corr} and \eqref{eq:VDFAP-CF-hess}, the correlation matrix of $\bs{N}$ is
\begin{align}
    \mathbb{E}\left[ \bs{N}\bs{N}^\tsp \right]
    =
    \frac{\lambda}{|u|} \mathbb{I}_{d}
    ,
\label{eq:VDFAP-corr-result}
\end{align}
\color{black}
which, upon taking the trace, yields
\begin{align}
\mathbb{E}[ \|\bs{N}\|^2 ]
=
\frac{\lambda d}{|u|}.
\label{eq:second-moment}
\end{align}
Another consequence of \eqref{eq:VDFAP-corr-result} is that components of
$\bs{N}\sim\text{VDFAP}^{(d)}(u,\lambda)$
are pairwise uncorrelated. However, they are \emph{not} independent, since the PDF \eqref{eq:dD-VDFAP} cannot be decomposed as the product of its marginals. 
\color{black}

\subsection{(Weak) Stability}
\label{subsec:stable}

\color{black}
In this subsection, we aim to prove a weak stability property for the VDFAP distributions, which will be used later to establish a lower bound on the VDFAP channel capacity.
\color{black}


\color{black}
Gaussian distributions, along with
Cauchy distributions 
and 
L{\'e}vy distributions, 
are known to be examples of \emph{stable} distributions \cite{farsad2015stable}. 
\color{black}
Roughly speaking, a distribution is stable if any linear combination of two independent random vectors from this distribution still follows the same distribution. That is, it is akin to the \emph{closure} property of vector spaces.
By definition, a distribution is called stable if and only if it 
possesses a specific form of CF \cite[Eq.~(2)]{fallahgoul2013multivariate}, which is not satisfied by \eqref{eq:VDFAP-CF-revised}.
\color{black}
Therefore, VDFAP distributions are \emph{not} stable under this definition. However, despite not being (strictly) stable, VDFAP distributions still exhibit certain ``weaker" stability properties.
\color{black}

\color{black}
One such \emph{weak stability property} that is instrumental in deriving a lower bound \textcolor{black}{(refer to the proof of Theorem 1)} for the capacity of VDFAP channels is encapsulated in the following lemma.

\color{black}
\begin{lma}[Weak Stability Property for VDFAP]
\label{lma:weak-stability}
If
$
\bs{N}_1\sim\text{VDFAP}^{(d)}(u,\lambda_1)
$ and 
$
\bs{N}_2 \sim\text{VDFAP}^{(d)}(u,\lambda_2)
$
are two independent VDFAP random vectors with the same normalized drift $u$, then their sum 
$
    \bs{N}_1+\bs{N}_2 
    \sim\text{VDFAP}^{(d)}(u,\lambda_1+\lambda_2).
$
\end{lma}
\color{black}

\begin{proof}
To prove this property, note that $\bs{N}_1$ and $\bs{N}_2$ are independent. Therefore, the CF of their sum is the product of their respective CFs \cite[Theorem 3.3.1]{lukacs1970characteristic}, i.e.,
$
\mathbb{E}[
    \exp(i\boldsymbol{\omega}^\tsp(\bs{N}_1+\bs{N}_2))]
    =
    \Phi^{(d)}_{\bs{N}_1}(\boldsymbol{\omega} ;u,\lambda_1)
    \Phi^{(d)}_{\bs{N}_2}(\boldsymbol{\omega} ;u,\lambda_2).
$
Using the CF formula \eqref{eq:VDFAP-CF-revised} for $\bs{N}_1$ and $\bs{N}_2$, we obtain
\begin{align}
\begin{split}
    &
    \mathbb{E}[
    \exp(i\boldsymbol{\omega}^\tsp(\bs{N}_1+\bs{N}_2))]
    \\
    =
    &
    \exp
            \left(
            -\lambda_1
            \left(
            \sqrt{\|\boldsymbol{\omega}\|^2+|u|^2}
            -|u|
            \right)
            \right)\\
    &\cdot
    \exp
            \left(
            -\lambda_2
            \left(
            \sqrt{\|\boldsymbol{\omega}\|^2+|u|^2}
            -|u|
            \right)
            \right)
    \\=&
    \exp
            \left(
            -\left(\lambda_1+\lambda_2\right)
            \left(
            \sqrt{\|\boldsymbol{\omega}\|^2+|u|^2}
            -|u|
            \right)
            \right) 
    .
\end{split}
\end{align}
Since the CF uniquely characterizes the distribution \cite[Theorem 3.1]{lukacs1970characteristic}, we can compare this result to the CF formula \eqref{eq:VDFAP-CF-revised} to conclude that $\bs{N}_1+\bs{N}_2 \sim \text{VDFAP}^{(d)}(u,\lambda_1+\lambda_2)$.
\end{proof}

\section{Bounds on the Capacity of VDFAP Channel}
\label{sec:bounds}
Borrowing the wisdom from parallel Gaussian channels
with a common power constraint\footnote{Although the noise components are assumed to be independent in the setting of parallel Gaussian channels, this is not the case for VDFAP channels when $d\geq 2$, as we argued at the end of Sec.~\ref{subsec:mean_and_cov}.} (see \cite[Section 10.4]{cover1999elements}),
we investigate the VDFAP channel capacity under
the second-moment constraint as follows:
\begin{align}
\begin{split}
    C 
    =     \sup_{p(\bs{X}):~\mathbb{E} \left[\|\bs{X}\|^2\right] \leq P}
    I(\bs{X};\bs{Y}),
\label{eq:def-cap-3D-VDFAP}
\end{split}
\end{align}
where the objective function $I(\bs{X};\bs{Y})$ is the \emph{mutual information} \cite{cover1999elements} between two $d$-dim random vectors $\bs{X}$ and $\bs{Y}$. The supremum is taken over all input 
distributions $p(\bs{X})$ satisfying
a constraint $\mathbb{E} \left[\|\bs{X}\|^2\right] \leq P$ we call the \emph{second-moment constraint}\footnote{Mathematically, this is identical to the \emph{common power constraint} for parallel Gaussian channels. But for FAP channels, $\mathbb{E} \left[\|\bs{X}\|^2\right]$, which has the unit of squared length, cannot be interpreted as power in the physical sense.}, where $P>0$ is an upper bound on the second moment
of the input random vector $\bs{X}$.
Note that the VDFAP channel output $\bs{Y}$ equals to $\bs{X}+\bs{N}$, where the noise $\bs{N}$ follows $\text{VDFAP}^{(d)}(u,\lambda)$ distribution, and is \emph{independent} of the input $\bs{X}$. The capacity appeared in \eqref{eq:def-cap-3D-VDFAP} depends on the dimension $d$ and the parameter triplet 
$(u,\lambda,P)$, so we sometimes write it as $C^{(d)}(u,\lambda,P)$ to stress these dependencies.


Applying the additive channel structure $\bs{Y}=\bs{X}+\bs{N}$ 
and the independence of $\bs{X}$ and $\bs{N}$, 
we have (assuming all differential entropies $h(\cdot)$ involved are well-defined and finite) $I(\bs{X};\bs{Y})=h(\bs{Y})-h(\bs{Y}|\bs{X})=h(\bs{Y})-h(\bs{N}|\bs{X})=h(\bs{Y})-h(\bs{N})$, and thus
\eqref{eq:def-cap-3D-VDFAP} can be simplified as:
\begin{align}
\begin{split}
    C
    =
     \sup_{p(\bs{X}):~\mathbb{E} \left[\|\bs{X}\|^2 \right] \leq P}
    h(\bs{X}+\bs{N}) 
    -
    h(\bs{N})
    ,
\label{eq:cap-3D-VDFAP}
\end{split}
\end{align}
where $h(\bs{W})$ denotes the differential entropy of a $d$-dim random vector $\bs{W}$ with support $S$ and PDF $f(\bs{w})$. Namely, 
$
    h(\bs{W})
    :=
    -
    \int_S
    f(\bs{w})
    \log
    f(\bs{w})
    ~\mathrm{d}\bs{w}
    ,
$
where $\log(\cdot)$ denotes the \emph{natural logarithm} function throughout this paper. Two things to note here:
\begin{itemize}
    \item[i)] 
    In this Sec.~\ref{sec:bounds} and Appendix~\ref{sec:appen-DE}, the differential entropy $h(\cdot)$, the capacity $C$ and its bounds are all measured in the unit of \emph{nats} for ease of expression. In Sec.~\ref{sec:numerical}, when plotting bounds on $C$ we use the unit of \emph{bits} for ease of interpretation. $1~\text{nats}=\log_2 e~\text{bits}\approx 1.44~\text{bits}$.
    \item[ii)] 
    Abusing the notation, we allow the argument of $h(\cdot)$ to be a random vector or the name of a distribution. For example, we write $h(\text{VDFAP}^{(d)}(u,\lambda))$ to denote the differential entropy of the $\text{VDFAP}^{(d)}(u,\lambda)$ distribution.
\end{itemize}

For $d=1$, to our best knowledge, no closed-form expression for the differential entropy of $\text{VDFAP}^{(1)}(u,\lambda)$ is known. However, for $d=2$, we have calculated a closed-form expression for differential entropy of a random vector $\bs{N}$ following $\text{VDFAP}^{(2)}(u,\lambda)$. The formula is
\begin{align}
\begin{split}
&    h(\text{VDFAP}^{(2)}(u,\lambda))
\\=~&
    \log \left( 2\pi e^3\right)
    + 2\log(\lambda)
    -
    \log (1+\lambda|u|)\\
    &-
    \lambda |u| e^{\lambda |u|}
    \big(
    e \cdot \text{Ei}(-1-\lambda |u|)
    -
    3 \cdot \text{Ei}(-\lambda |u|)
    \big)
    ,
\label{eq:diff-ent-3D-VDFAP}
\end{split}
\end{align}
where the exponential integral function $\text{Ei}(\cdot)$ is defined as \textcolor{black}{\cite[Section~8.21]{gradshteyn2014table}}:
\begin{align}
    \text{Ei}(x)
    =
    - \int_{-x}^\infty \frac{e^{-t}}{t} 
    \diff t
    , \text{\ for\ } x < 0
    .
\label{eq:def-exp-int}
\end{align}
The detailed calculation of \eqref{eq:diff-ent-3D-VDFAP} \textcolor{black}{is presented in Appendix~\ref{sec:appen-DE}.}


\color{black}
Next, we aim to establish two theorems characterizing the lower and upper bounds for the VDFAP channel capacity $C$. By using the derived closed-form expression 
\eqref{eq:diff-ent-3D-VDFAP}, we further derive closed-form lower and upper bounds for $C$ in 3D space ($d=2$).
\color{black}

\subsection{Lower Bound on Capacity}

Our strategy is to apply the weak stability property of VDFAP distributions proved in Section~\ref{subsec:stable} to obtain a lower bound 
of the VDFAP channel capacity $C=C^{(d)}(u,\lambda,P)$. The result is stated in the following Theorem.
\color{black}
\begin{thm}[Lower Bound on $C$]
    \label{thm:lower-bound-general}
    The capacity \(C^{(d)}(u,\lambda,P)\) is lower bounded as follows:
    \begin{align}
        \begin{split}
        C^{(d)}(u,\lambda,P)
        \geq\ &
        \sup_{0 < \lambda' \leq |u| P/d}
        h\left(
        \text{VDFAP}^{(d)}
        \left(
        u, \lambda + \lambda'
        \right)
        \right)
        \\
        &-
        h\left(
        \text{VDFAP}^{(d)}
        \left(
        u, \lambda
        \right)
        \right)
        .
        \label{eq:lower-bound-general}
    \end{split}
\end{align}
\end{thm}
\color{black}

\begin{proof}
    A general strategy 
    to obtain lower bounds on the capacity
    \eqref{eq:cap-3D-VDFAP} of additive channels is to
    pick 
    a family of
    parametrized
    input distributions 
    $p_\theta(\bs{X})$ satisfying the 
    specified constraint(s), calculate the corresponding 
    $h(\bs{X}+\bs{N})-h(\bs{N})$ 
    for each parameter $\theta$, and then optimize the result via taking supremum
    over $\theta$.

    Applying this general strategy to VDFAP channels under the second-moment constraint while having the \emph{weak stability} Lemma (proved in Sec.~\ref{lma:weak-stability}) in mind, we select the family of input distributions to be VDFAP with the \emph{same} $u$ as the noise but a generally different $\lambda'$ such that the second-moment constraint is satisfied. That is,
    $\bs{X}\sim\text{VDFAP}^{(d)}(u,\lambda')$ for 
    some strictly positive $\lambda'\leq |u|P/d$.
    By \eqref{eq:second-moment}, we have that 
    $
        \mathbb{E}\left[\|\bs{X}\|^2\right]
        =
        \frac{\lambda'd}{|u|}
        \leq
        \frac{(|u|P/d)d}{|u|}
        =
        P
        ,
    $
    so $\bs{X}$ satisfies the 
    second-moment constraint.
    Also, by \emph{weak stability} Lemma,
    $\bs{Y}=\bs{X}+\bs{N}$ follows $\text{VDFAP}^{(d)}(u,\lambda+\lambda')$. Then, a lower bound that holds for any $0<\lambda'\leq 
    |u|P/d$ is
    \begin{align}
        \begin{split}
        &C^{(d)}(u,\lambda,\Sigma)\\
        \geq~&h(\bs{X}+\bs{N}) - h(\bs{N})\\
        =~&
        h
        \left(
        \text{VDFAP}^{(d)}(u,\lambda+\lambda')
        \right)
        -
        h
        \left(
        \text{VDFAP}^{(d)}(u,\lambda)
        \right)
        ,
        \label{eq:lower-bound-most-general}
        \end{split}
    \end{align}
    where the inequality follows directly from \eqref{eq:cap-3D-VDFAP}. 
    Taking the supremum over all admissible $\lambda'$
    proves \eqref{eq:lower-bound-general}.
\end{proof}

\color{black}
\begin{cor}[Lower Bound on $C$ for \(d=2\)]
    \label{thm:lower-bound}
    A closed-form expression of the lower bound given in Theorem~\ref{thm:lower-bound-general} for $d=2$ is
\begin{align}
\begin{split}
    C^{(2)}(u,\lambda,P)           
    &\geq
    2 \log \Big( 1 + \tfrac{P|u|}{2\lambda} \Big)
    -
    \log \left( 1 + \tfrac{P|u|^2}{2(1+\lambda |u|)} \right)
    \\
    &~~
    +
    \lambda |u| e^{\lambda |u|}
    \big(
    e \cdot \text{Ei}(-1-\lambda |u|)
    -
    3 \cdot \text{Ei}(-\lambda |u|)
    \big)
    \\
    &~~
    -
    \Big(
    \lambda+\tfrac{P|u|}{2}
    \Big)
    |u| \exp\left\{\left(
    \lambda+\tfrac{P|u|}{2}
    \right) |u|\right\}
    \\
    &~~
    \times
    \bigg\lbrace
    e \cdot 
    \text{Ei}
    \Big(
    -1-
    \big(
    \lambda+\tfrac{P|u|}{2}
    \big) 
    |u|
    \Big)
    \\
    &~~~~~~
    -
    3 \cdot 
    \text{Ei}
    \Big(
    -
    \big(
    \lambda+\tfrac{P|u|}{2}
    \big) 
    |u|
    \Big)
    \bigg\rbrace
    \\
    &>
    0
    .
\label{eq:lower-bound}
\end{split}
\end{align}
\end{cor}
\color{black}

\begin{proof}

    To obtain the closed-form expression for the lower bound in 
    \eqref{eq:lower-bound-general} for $d=2$,
    we introduce an ancillary function $h_0: \R^+ \to \R$ (where $\R^+=\{x\in\mathbb{R}:x>0\}$) defined by
    \begin{align}
    \begin{split}
        h_0(s)
        :=~&
        2 \log (s) 
        - \log (1+s) \\
        &- s e^s 
        \big(
        e \cdot \text{Ei}(-1-s)
        - 3 \cdot \text{Ei}(-s)
        \big)
        ,
    \end{split}
    \label{eq:def-h0}
    \end{align}
    so that the differential entropy \eqref{eq:diff-ent-3D-VDFAP} of a random vector $\bs{Z}\sim\text{VDFAP}^{(2)}(u,\lambda'')$ can be expressed as
    $
        h(\mathbf{Z})
        =
        h_0(|u|\lambda'') + \log (2\pi e^3) - 2 \log (|u|)
        .
    $
    Therefore, the lower bound \eqref{eq:lower-bound-general} can be expressed via the function $h_0(\cdot)$ as 
    \begin{align}
        C^{(2)}(u,\lambda,P)
        \geq
        \sup_{0<\lambda'\leq |u|P/2}
        h_0(|u|(\lambda+\lambda'))
        -
        h_0(|u|\lambda)
        .
    \label{eq:lower-bound-3}
    \end{align}
    \color{black}
    It is shown \textcolor{black}{in
    Appendix~\ref{sec:appen-SI}}
    that $h_0(\cdot)$ is strictly monotonically increasing.
    \color{black}
    Therefore, \eqref{eq:lower-bound-3} boils down to plugging the maximum $\lambda'$, i.e., 
    $|u|P/2$, into 
    \eqref{eq:lower-bound-3}: 
    \begin{align}
        \begin{split}
        C^{(2)}(u,\lambda,P)
        &\geq
        h_0
        \bigg(
        |u|
        \Big(
        \lambda
        +
        |u| P/2
        \Big)
        \bigg)
        -
        h_0(|u|\lambda)>0.
        \end{split}    
    \label{eq:lower-bound-2}
    \end{align}
    The strict positivity of the lower bound in \eqref{eq:lower-bound-2} 
    also
    follows from the 
    strictly monotonically increasing property
    of $h_0(\cdot)$ and  
    the strict positivity of $|u|$ and $P$.
    \textcolor{black}{Also note that the input distribution achieving this lower bound is exactly $\text{VDFAP}(u,|u|P/2)$, according to the proof of Theorem~\ref{thm:lower-bound-general}.}
    
    Last but not least,
    the closed-form lower bound in \eqref{eq:lower-bound} can be obtained by directly plugging the expression \eqref{eq:def-h0} of $h_0$ into the lower bound in \eqref{eq:lower-bound-2}. 
    The strict positivity of the lower bound in \eqref{eq:lower-bound} follows directly.
\end{proof}

\subsection{Upper Bound on Capacity}

It is well known 
\cite[Theorem~2.7, (2.20)]{polyanskiy2022book}
in information theory
that 
under a prescribed 
upper bound $Q>0$ on the second moment $\mathbb{E}[\|\bs{Y}\|^2]$ 
of a continuous random vector $\bs{Y}\in\mathbb{R}^d$,
the differential entropy $h(\bs{Y})$ 
is maximized by the multivariate Gaussian 
$\mathcal{N}\left(\bs{0},\frac{Q}{d}\mathbb{I}_d\right)$.
That is,
\begin{align}
    \max_{q(\bs{Y}):\mathbb{E}[\|\bs{Y}\|^2]\leq Q}
    h(\bs{Y})
    =
    h\left(\mathcal{N}\left(\bs{0},\tfrac{Q}{d}\mathbb{I}_d\right)\right)
    =
    \tfrac{d}{2} \log\left( 2\pi e \tfrac{Q}{d} \right)
    ,
    \label{eq:max-cap-Gauss}
\end{align}
where $q(\bs{Y})$ denotes the distribution of $\bs{Y}$.
Our strategy is to apply this 
well-known fact to establish an upper bound 
for the VDFAP channel capacity $C=C^{(d)}(u,\lambda,P)$. The result is stated in the following theorem.
\color{black}
\begin{thm}[Upper Bound on $C$]
    \label{thm:upper-bound-general}
    The capacity \(C^{(d)}(u,\lambda,P)\) is upper bounded as follows:
\begin{align}
    \begin{split}
    C^{(d)}(u,\lambda,P)
    \leq~&
            \tfrac{d}{2}
                \log
                \left(
                2\pi e
                \left(
                \tfrac{P}{d}
                +
                \tfrac{\lambda}{|u|}
                \right)
                \right)\\
            &-
            h
            \left( 
            \text{VDFAP}^{(d)}(u,\lambda) 
            \right)
            .            
    \end{split}
    \label{eq:upper-bound-most-general}
\end{align}
\end{thm}

\color{black}
\color{black}

    \begin{proof}
        Consider \eqref{eq:cap-3D-VDFAP} as the maximization problem
        \begin{align}
            C
            =
            \sup_{ p(\bs{X}) \in \mc{F}_{P} }
            h(\bs{X}+\bs{N}) - h(\bs{N})
            \label{eq:cap-max-input}
        \end{align}
        where the constraint set 
        $\mc{F}_{P}$ 
        is the family of input distributions
        $
            \mc{F}_{P}
            =
            \left\lbrace
            p(\bs{X}):~
            \mathbb{E}\left[ 
            \|\bs{X} \|^2 
            \right]
            \leq P
            \right\rbrace
            .
        $
        For the VDFAP channel, we can calculate the 
        second moment 
        of $\bs{Y}=\bs{X}+\bs{N}$
        as
    \begin{align}
        \begin{split}
            \mathbb{E}[\| \bs{Y} \|^2]
            &=
            \mathbb{E}[\| \bs{X} + \bs{N} \|^2]
            \\&=
            \mathbb{E}[\| \bs{X} \| ^2 + 2 \bs{X}^\tsp \bs{N}  + \| \bs{N} \| ^2]
            \\&=
            \mathbb{E}[\| \bs{X} \| ^2] + 2 \mathbb{E}[ \bs{X}^\tsp \bs{N} ] + \mathbb{E}[\| \bs{N} \| ^2]
            \\&\overset{(a)}{=}
            \mathbb{E}[\| \bs{X} \| ^2] + 2 \mathbb{E}[ \bs{X}]^\tsp \mathbb{E}[ \bs{N}] + \mathbb{E}[\| \bs{N} \| ^2]
            \\&\overset{(b)}{=}
            \mathbb{E}[\| \bs{X} \| ^2] + \mathbb{E}[\| \bs{N} \| ^2]
            \\&\overset{(c)}{=}
            \mathbb{E}[\| \bs{X} \| ^2] + \tfrac{\lambda d}{|u|}
            ,
        \label{eq:output-power-constr}
        \end{split}
    \end{align}
        where (a) is due to the independence of $\bs{X}$ and $\bs{N}$,
        (b) is due to $\mathbb{E}[ \bs{N} ]=\boldsymbol{0}$ and (c) is due to \eqref{eq:second-moment}.
        Defining $Q=P+\lambda d/|u|$ and the corresponding family of second-moment constrained output distributions as
        \(
            \mathcal{G}_{Q}
            =
            \left\lbrace
            q(\bs{Y}):~
            \mathbb{E}\left[ 
            \|\bs{Y}\|^2 
            \right]
            \leq
            Q
            \right\rbrace
            ,
        \)
        we have that \eqref{eq:output-power-constr} implies: for any $P>0$,
        \begin{align}
            \begin{split}
            &
            p(\bs{X}) \in \mathcal{F}_P
            \text{ and } \bs{Y}=\bs{X}+\bs{N}
            ~(\bs{X},\bs{N} \text{ independent}) 
            \\
            \Longrightarrow~&
            q(\bs{Y}) \in \mathcal{G}_{Q}
            .
            \end{split}
            \label{eq:cap-relax}
        \end{align}
        In other words, if there is an input distribution $p(\bs{X})$ satisfying the second-moment constraint $P$, then the corresponding output $\bs{Y}=\bs{X}+\bs{N}$ (with $\bs{X}, \bs{N}$ independent) has a distribution $q(\bs{Y})$ satisfying the second-moment constraint $Q$. However, the converse statement does \emph{not} hold: even if a desired output distribution $q(\bs{Y})$ satisfies a second-moment constraint, there is \emph{no} guarantee that a corresponding input distribution $p(\bs{X})$ should exist such that $\bs{Y}=\bs{X}+\bs{N}$. Hence, the following maximization problem \begin{align}
            C'
            =
            \sup_{ q(\bs{Y}) \in \mc{G}_{Q} }
            h(\bs{Y}) - h(\bs{N})
            \label{eq:cap-max-output}
        \end{align}
        is a \emph{relaxation} of the original problem \eqref{eq:cap-max-input}, 
        and because of this relaxation, we have $C\leq C'$.
        To conclude, 
        from \eqref{eq:max-cap-Gauss} 
        and
        \eqref{eq:cap-max-output} we have 
        \begin{align}
            C
            \leq
            C'
            =
            \tfrac{d}{2} \log\left( 2\pi e \tfrac{Q}{d} \right)
            -
            h
            \left( 
            \text{VDFAP}^{(d)}(u,\lambda) 
            \right)
            ,
        \label{eq:upper-bound-2}
        \end{align}
        and \eqref{eq:upper-bound-most-general} follows upon replacing $Q$ by its definition.
\end{proof}

\color{black}
\begin{cor}[Upper Bound on $C$ for \(d=2\)]
    \label{thm:upper-bound}
    \begin{align}
    \begin{split}
        C^{(2)}(u,\lambda,P)
        &
        \leq 
        \log 
        \left(
        \tfrac{P}{2\lambda^2}
        +
        \tfrac{1}{\lambda |u|} \right)
        +
        \log (1+\lambda|u|)
        -
        2        
        \\
        &
        +
        \lambda |u| e^{\lambda |u|}
        \left(
        e \cdot \text{Ei}(-1-\lambda |u|)
        -
        3 \cdot \text{Ei}(-\lambda |u|)
        \right)
        .
    \label{eq:upper-bound}
    \end{split}
    \end{align}
\end{cor}
\color{black}
\begin{proof}
    To get a closed-form upper bound when $d=2$, we plug the differential entropy expression \eqref{eq:diff-ent-3D-VDFAP} into \eqref{eq:upper-bound-most-general} and simplify to get \eqref{eq:upper-bound}.
\end{proof}

\color{black}

\begin{figure}[ht]
    \centering
    \begin{subfigure}{\columnwidth} 
        \includegraphics[width=0.9\linewidth]{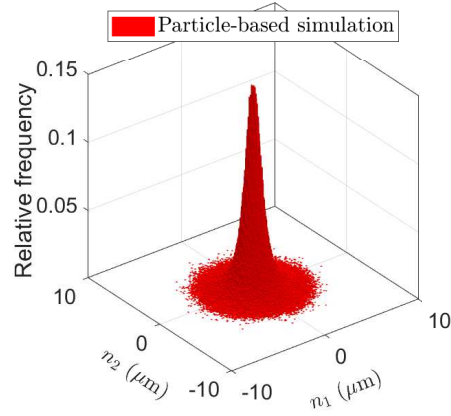}
        \caption{
        Particle-based simulated FAP density on a planar receiver when drift is zero. \textcolor{black}{The vertical height represents the relative frequency, estimated via the Monte Carlo method.
        }
        }
        \label{fig:planarRX_particle}
    \end{subfigure}

    \begin{subfigure}{\columnwidth} 
        \includegraphics[width=0.9\linewidth]{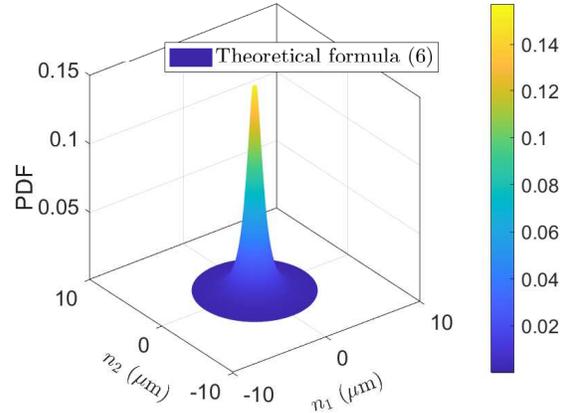}
        \caption{
        \textcolor{black}{
        Theoretical FAP density 
        on a planar receiver when drift is zero.
        }
        }
        \label{fig:planarRX_numerical}
    \end{subfigure}
    \caption{
    \textcolor{black}{
    Comparison of particle-based simulated and theoretical FAP densities on a planar receiver (i.e., $d=2$) for zero drift $\mathbf{u}=[0,0,0]$.
    }
    }
    \label{fig:planarRX_both}
\end{figure}

\begin{figure}[ht]
    \centering
    \begin{subfigure}{\columnwidth} 
        \includegraphics[width=0.9\linewidth]{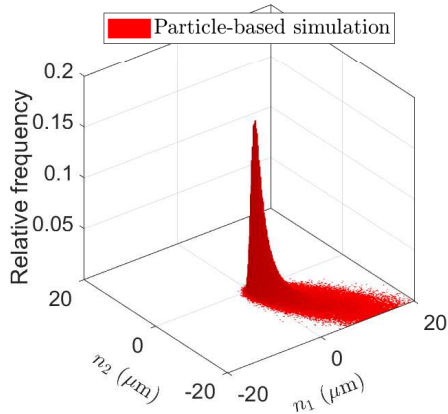}
        \caption{
        Particle-based simulated FAP density on a planar receiver when 
        $\mathbf{u}$ is $[2, -3, -1]$ $(\mu \text{m}^{-1})$.
        \textcolor{black}{The vertical height represents the relative frequency, estimated via the Monte Carlo method.
        }
        }
        \label{fig:planarRX_drift_particle}
    \end{subfigure}

    \begin{subfigure}{\columnwidth} 
        \includegraphics[width=0.9\linewidth]{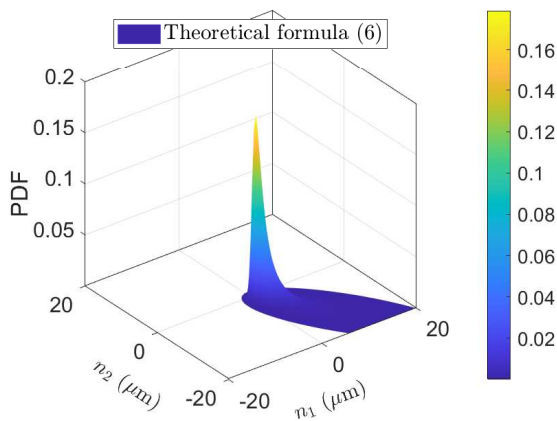}
        \caption{
        \textcolor{black}{
        Theoretical FAP density 
        on a planar receiver when 
        $\mathbf{u}$ is $[2, -3, -1]$ $(\mu \text{m}^{-1})$.
        }
        }
         \label{fig:planarRX_drift_numerical}
    \end{subfigure}
    \caption{
    \textcolor{black}{
    Comparison of particle-based simulated and theoretical FAP densities on a planar receiver (i.e., $d=2$) for a case of non-zero normalized drift $\mathbf{u}=[2,-3,-1]~(\mu\mathrm{m}^{-1})$.
    }
    }
    \label{fig:planarRX_drift_both}
\end{figure}

\section{Numerical Results} \label{sec:numerical}

\subsection{Particle-based Verification of the Derived FAP Density Formulas}

In the 3D scenario ($d=2$) depicted in Fig.~\ref{fig:3Da}, where molecules are released from a point source at $[0,0,1]~(\mu\mathrm{m})$ (i.e., the Tx-Rx distance $\lambda = 1~(\mu\mathrm{m})$), we carried out particle-based simulations (PBS) and numerical evaluations of the theoretical FAP densities as per 
\eqref{eq:dD-FAP}.
We examined two different normalized drift vectors, one being zero $\mathbf{u}=[0,0,0]$ and the other non-zero $\mathbf{u}=[2,-3,-1]~(\mu\mathrm{m}^{-1})$. The resulting PDFs of position deviation, referred to as FAP noise $[n_1,n_2]$, are displayed in Fig.~\ref{fig:planarRX_both} and Fig.~\ref{fig:planarRX_drift_both}, respectively.

For the PBS, we utilized the Monte Carlo method as detailed in \cite[Eq. (9)-(11)]{farsad2016comprehensive} to simulate $M=10^6$ independent trajectories of dissolved Ethanol molecules diffusing in water with a diffusion coefficient $D \approx 840~(\mu\mathrm{m}^2/\mathrm{sec})$ and a time step of $10^{-6}~(\mathrm{sec})$.\footnote{To avoid possible confusion with the noise $N$, we use $M$ to denote the number of particles.} The first hitting position of each molecule was approximated by recording the first two coordinates when the third coordinate first becomes negative, indicating crossing the receiver plane at $x_3=0$. The derived FAP density estimates are presented in Fig.~\ref{fig:planarRX_particle} and Fig.~\ref{fig:planarRX_drift_particle}, corresponding to each drift scenario. The theoretical predictions were computed using the derived FAP density formula \eqref{eq:dD-FAP}.
In producing these figures, we truncated values below $10^{-3}$ to emphasize the approximate support of the distribution, although theoretically, the FAP distributions could be supported on the whole $\mathbb{R}^2$ plane.

The PBS results in the no-drift scenario (Fig.~\ref{fig:planarRX_particle}) show a symmetrical distribution of particles around the origin, aligning with our theoretical predictions (Fig.~\ref{fig:planarRX_numerical}). In the scenario with drift, the particle distribution skews in the direction of the parallel normalized drift $\bs{u}_\text{par}=[2,-3]~(\mu\mathrm{m}^{-1})$, as seen in Fig.~\ref{fig:planarRX_drift_particle}, which is consistent with our theoretical expectations (Fig.~\ref{fig:planarRX_drift_numerical}).

Figures~\ref{fig:planarRX_both} and \ref{fig:planarRX_drift_both} demonstrate a strong alignment between our theoretical predictions and the PBS results, with a simulation error in density estimate on the order of $1/\sqrt{M}=10^{-3}$. This agreement provides strong evidence for the accuracy and reliability of our derived FAP density function. 

\subsection{Graphical Illustration of the Trends of Capacity Bounds}

\begin{figure}
    \centering
    \begin{subfigure}{0.9\columnwidth}
        \centering
        \includegraphics[width=\linewidth]{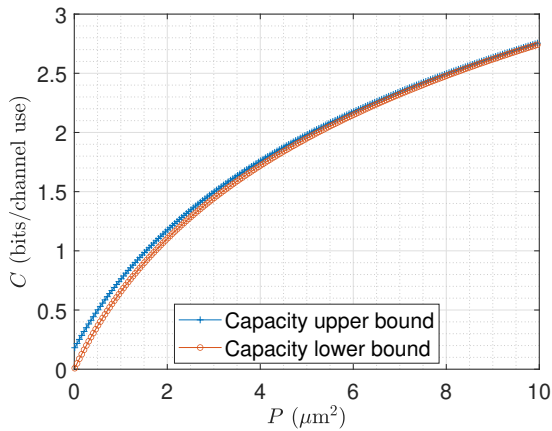}
        \caption{
        \textcolor{black}{
        Capacity bounds versus second-moment constraint \( P \) with $\lambda=1~(\mu\mathrm
        {m})$ and $u=-1~(\mu\mathrm{m}^{-1})$ fixed.
        }
        }
        \label{fig:cap-P}
    \end{subfigure}
    \hfill
    \begin{subfigure}{0.9\columnwidth}
        \centering
        \includegraphics[width=\linewidth]{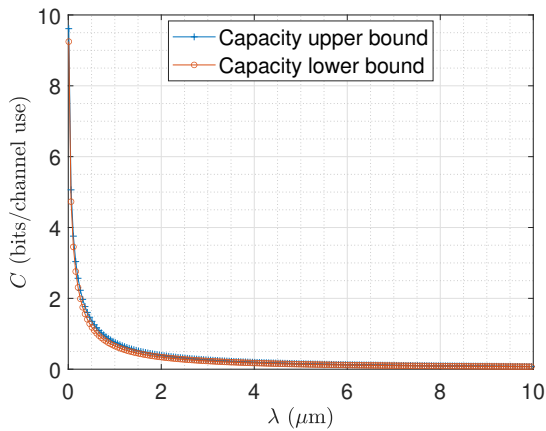}
        \caption{
        \textcolor{black}{
        Capacity bounds versus transmitter-receiver distance \( \lambda \) with $P=1~(\mu\mathrm
        {m}^2)$ and $u=-1~(\mu\mathrm{m}^{-1})$ fixed.
        }
        }
        \label{fig:cap-lambda}
    \end{subfigure}
    \hfill
    \begin{subfigure}{0.9\columnwidth}
        \centering
        \includegraphics[width=\linewidth]{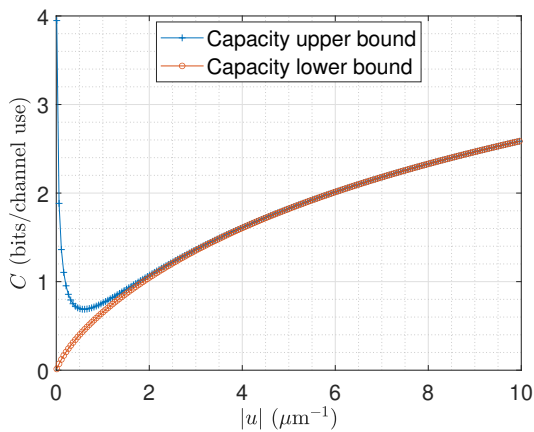}
        \caption{
        \textcolor{black}{
        Capacity bounds versus absolute normalized drift \( |u| \) with $P=1~(\mu\mathrm
        {m}^2)$ and $\lambda=1~(\mu\mathrm
        {m})$ fixed.
        }
        }
        \label{fig:cap-u}
    \end{subfigure}
    \caption{
    \textcolor{black}{
    Numerical evaluations of derived lower bound \eqref{eq:lower-bound} and upper bound \eqref{eq:upper-bound} on the capacity of VDFAP channels when \( d=2 \), illustrating the influence of \( P \), \( \lambda \), and \( u \) on the trends of capacity. 
    }
    }
    \label{fig:three-graphs}
\end{figure}

This subsection presents numerical evaluations illustrating the impact of key parameters on the capacity bounds derived in \eqref{eq:lower-bound} and \eqref{eq:upper-bound} for a 3D scenario ($d=2$). In Fig.~\ref{fig:three-graphs}, we choose the operating point $[P=1~(\mu\mathrm
        {m}^2), \lambda=1~(\mu\mathrm
        {m}), u=-1~(\mu\mathrm{m}^{-1})]$ and vary one parameter while keeping the other two fixed, to investigate the influence from each key parameter alone while the others are held constant.

In Fig.~\ref{fig:cap-P}, we demonstrate how capacity increases with the 
input second-moment constraint $P$
, when $\lambda=1~(\mu\mathrm{m})$ and $u=-1~(\mu\mathrm{m}^{-1})$ are fixed.
From the figure, the trend is logarithmic (also can be seen from the upper bound formula \eqref{eq:upper-bound}), similar to the case for parallel Gaussian channels (see \cite[Section 10.4]{cover1999elements}). Also, it can be seen that the gap between upper and lower bounds closes as $P$ grows, and the maximum gap is roughly $0.2$ (bits/channel use) for the range $P\in(0,10]~(\mu\mathrm{m}^2)$.

Fig.~\ref{fig:cap-lambda} illustrates the inverse relationship between capacity and the 
Tx-Rx distance $\lambda$, when $P=1~(\mu\mathrm{m}^2)$ and $u=-1~(\mu\mathrm{m}^{-1})$ are fixed. This inverse relationship is intuitive: the farther the Tx-Rx distance, the ``worse" the FAP channel is. 
It is also evident that the gap between upper and lower bounds closes as $\lambda$ grows, and the largest gap is less than $0.5$ (bits/channel use) for the range $\lambda\in(0,10]~(\mu\mathrm{m})$.

Finally,
Fig.~\ref{fig:cap-u} highlights the effect of 
normalized drift $u$ on capacity, when $P=1~(\mu\mathrm{m}^2)$ and $\lambda=1~(\mu\mathrm{m})$ are fixed.
It makes intuitive sense that when $|u|$ is large, i.e., when the relative strength of drift over diffusion is large, the VDFAP channel is more capable, as it takes a shorter duration to travel from the Tx to the Rx, and the diffusion effect, which introduces positional uncertainty during the same duration, is less pronounced.
From the figure, it is apparent that for $|u|$ larger than roughly $0.5~(\mu\mathrm{m}^{-1})$, the gap between upper and lower bounds closes as $|u|$ grows, and the maximum gap is roughly $0.3$ (bits/channel use) for this range of $|u|$. However, for $|u|$ smaller than $0.5~(\mu\mathrm{m}^{-1})$, it can be observed that the effectiveness of the upper bound deteriorates as $|u|\to 0$. In fact, from \eqref{eq:upper-bound} we see directly that the upper bound goes to $+\infty$ as $|u|\to 0$ while keeping $P$ and $\lambda$ fixed. This makes sense since the upper bound is obtained by ``matching'' the output second moment to a multivariate Gaussian. However, as $|u|\to 0$, the second moment of VDFAP noise $\bs{N}$ approaches $+\infty$, and by \eqref{eq:output-power-constr} so does the second moment of the multivariate Gaussian. 
\textcolor{black}{This raises the interesting question of whether the second-moment constraint is suitable when the VDFAP noise distribution approaches a multivariate Cauchy distribution (known for having an infinite second moment) as $|u|\to 0$. This topic is addressed in a separate work \cite{lee2022capacity} by some of the authors.}

These figures elucidate the comprehensiveness and tightness of the derived bounds, underscoring how the upper and lower bounds diverge or converge under various key parameters. This visual representation aids in understanding the nuanced effects of these parameters on the derived capacity bounds for VDFAP channels.

\color{black}
\section{Conclusion and Future Work} \label{sec:conc}

\subsection{Summary and Conclusion}
In this paper, we have developed a mathematical rigorous model for MC systems that harnesses FAP as a mode of information transmission. We tackled two fundamental challenges: i) characterizing the FAP density and ii) establishing capacity bounds for vertically-drifted FAP channels.

Our approach to characterizing FAP density in MC systems with a fully-absorbing receiver is based on bridging macroscopic PDE models with microscopic SDE models, resulting in a concise and versatile expression, as presented in Eq.~\eqref{eq:main}. This formula, which connects FAP density to elliptic-type Green's function, is notable for its broad applicability across any spatial dimensions, any drift directions, and various receiver geometries.

The practicality and accuracy of our approach were demonstrated through specific case studies involving various receiver geometries, including 2D linear, 3D planar, and 3D spherical receivers 
(as presented in Appendix~\ref{sec:case}).
These cases exemplify the versatility of our model. Further validation of our model's accuracy was achieved through particle-based simulations (in Section VI-A), affirming the precision of our derived formulas.

On the other hand, armed with explicit expressions for the FAP density, we have successfully established closed-form upper and lower bounds (Theorem 1 and Theorem 2) on the capacity of vertically-drifted FAP channels under a second-moment constraint. This represents a significant advancement in our understanding of the fundamental information transmission rate for FAP channels in MC systems.

\color{black}
\subsection{
Potential Applications and Future Work
}

The development of FAP channels in molecular communication systems is not only a theoretical advance but also holds significant practical implications.
Our research has particularly vital implications for Molecular MIMO \cite{koo2016molecular} and Molecular Index Modulation (Molecular-IM) systems.


\color{black}
In the field of Molecular-IM systems, as detailed in \cite{gursoy2019pulse} and \cite{gursoy2020molecular}, there has been a widespread adoption of ``Rx box" models and Uniform Circular Array (UCA) topologies, which are illustrated in \cite[Fig.~1]{gursoy2020molecular}. These models typically utilize spherical-shaped Rx antennas. Nonetheless, there are considerable practical difficulties in implementing perfectly spherical antennas, particularly when integrating them into flat surfaces, owing to the limited contact area they provide. This preference for spherical shapes stems from the foundational research by Yilmaz et al. \cite{yilmaz2014three} in 2014, which introduced a \emph{point-to-sphere} density formula. This formula has become a cornerstone in most contemporary Molecular MIMO related research, influencing studies like those conducted by 
Sabu et al. \cite{sabu20203}
and Kwak et al. \cite{kwak2020two}.
These investigations are instrumental in determining the proportion of MMs received by each antenna in both Molecular MIMO and Molecular-IM systems.
However, this approach seems to be constrained to scenarios where the receivers or antennas have a spherical configuration.

Our significant advancement in this area is the development of a \emph{point-to-plane} position transition model, as depicted in Fig.~\ref{fig:3Da}. 
\textcolor{black}{
This model enables the accommodation of various antenna geometries\footnote{
\textcolor{black}{
Recently, Brand et al. also discussed the scenario of point Tx and cylindrical-shaped Rx in \cite{brand2023area}.
}}, such as an ellipsoid-shaped antenna (which has not yet been explored in the literature).} 
By utilizing our point-to-plane formula and applying integration, we can approximate a curved surface with numerous small planes. This approach offers a promising solution for different antenna shapes.
Additionally, our model innovatively incorporates a non-zero background drift, a feature not addressed in the point-to-sphere formula proposed by Yilmaz et al. \cite{yilmaz2014three}. This inclusion marks a notable improvement over the existing methodologies.
We believe our new model represents a crucial step towards resolving the complexities associated with more general antenna geometries in Molecular MIMO and IM systems. 
\textcolor{black}{Finally, 
note that a very recent work \cite{vakilipoor2023heuristic} proposed a heuristic analytical model (utilizing negative particles) to describe the impulse response of a Molecular SIMO (i.e., Single-Input Multiple-Output) system with multiple Fully-Absorbing (FA) receivers. 
An interesting future direction is to integrate our theoretical model with this heuristic model, and compare their predictive performances with PBS.
}

\color{black}

\appendices
\section{
\color{black}
Derivation of the Characteristic Function
\color{black}
} \label{sec:appen-CF}



We apply the Fourier transform (FT) pair relationship between the PDF and CF of Student's $t$-distribution to obtain an integral representation for \eqref{eq:dD-VDFAP}. 
Specifically, a random variable $N$ following the Student's $t$-distribution of $\nu$ degrees of freedom ($\nu>0$) has the PDF \cite{gaunt2021simple}:
\begin{align}
    f^{(\nu)}_N(n)
    &=
    \frac{\Gamma\left(\frac{\nu+1}{2}\right)}{\sqrt{\nu \pi} \Gamma\left(\frac{\nu}{2}\right)}
    \left(1+\tfrac{n^2}{\nu}\right)^{-\tfrac{\nu+1}{2}}
    ,\text{\ for\ } 
    n \in \mathbb{R},
\end{align}
(where $\Gamma(\cdot)$ is the gamma function) and the CF \cite{gaunt2021simple}:
\begin{align}
    \mathbb{E}[e^{i\omega N}]
    =
    \frac{K_{\nu/2}(\sqrt{\nu}|\omega|) \cdot(\sqrt{\nu}|\omega|)^{\tfrac{\nu}{2}}}{\Gamma\left(\frac{\nu}{2}\right) 2^{\frac{\nu-2}{2}}}
    ,\text{\ for\ } 
    \omega \in \mathbb{R}
.\end{align}
The FT pair relationship yields that 
\begin{align} \label{eq:Stu-FT-pair}
    K_{\nu/2}(\sqrt{\nu}|\omega|)
    &=
    \frac{2^{\frac{\nu-2}{2}}\Gamma\left(\frac{\nu+1}{2}\right)}{(\sqrt{\nu}|\omega|)^{\frac{\nu}{2}}\sqrt{\nu \pi}}
    \int_{\mathbb{R}}
    \left(1+\tfrac{n^2}{\nu}\right)^{-\tfrac{\nu+1}{2}}
    e^{i\omega n}
    ~\mathrm{d}n
    ,\end{align}
for $\omega\in\mathbb{R}\setminus\{0\}$.
Fix any $s>0$. Applying the substitutions $\tilde{n}=(s/\sqrt{\nu})n$ and $\tilde{\omega}=(\sqrt{\nu}/s)\omega$ to \eqref{eq:Stu-FT-pair}, we have 
\begin{align} \label{eq:K-Int-Rep}
    \frac{K_{\nu/2}(s|\tilde{\omega}|)}{(s|\tilde{\omega}|)^{\frac{\nu}{2}}}
    &=
    \frac{2^{\frac{\nu-2}{2}}\Gamma\left(\frac{\nu+1}{2}\right)}{\sqrt{\pi}|\tilde{\omega}|^{\nu}}
    \int_{\mathbb{R}}
    \left(s^2+\tilde{n}^2\right)^{-\tfrac{\nu+1}{2}}
    e^{i\tilde{\omega}\tilde{n}}
    ~\mathrm{d}\tilde{n}
    ,
\end{align}
for $\tilde{\omega}\in\mathbb{R}\setminus\{0\}$.
Setting $\nu=d+1$, $s=(\norm{\bs{n}}^2+\lambda^2)^{1/2}$ and $\tilde{\omega}=\vert u \vert$ in \eqref{eq:K-Int-Rep}, and then plugging into \eqref{eq:dD-VDFAP} yields
\begin{align}
\begin{split}
    f^{(d)}_{\bs{N}}(\bs{n};u,\lambda)
    &=
    \frac{\Gamma\left(\frac{d+2}{2}\right)}{\pi^{\frac{d+2}{2}}} \lambda e^{\lambda\vert u \vert} 
    \\
    &\quad
    \cdot
    \int_{\mathbb{R}} \left(\norm{\bs{n}}^2+\tilde{n}^2+\lambda^2\right)^{-\tfrac{d+2}{2}}
    e^{i\tilde{\omega}\tilde{n}}
    ~\mathrm{d}\tilde{n}. 
\label{eq:dD-VDFAP-PDF-int-rep}
\end{split}
\end{align}
Thus for the $d$-dim VDFAP distribution, its CF \eqref{eq:CF-def}:
\begin{align}
    \Phi^{(d)}_{\bs{N}}(\boldsymbol{\omega};u,\lambda)
    &=
    \int_{\mathbb{R}^{d}} f^{(d)}_{\bs{N}}(\bs{n};u,\lambda)
    ~e^{i\boldsymbol{\omega}^\intercal \bs{n}} 
    ~\mathrm{d}\bs{n}
\label{eq:dD-VDFAP-CF-int-rep}
\end{align}
can be obtained by plugging \eqref{eq:dD-VDFAP-PDF-int-rep} into \eqref{eq:dD-VDFAP-CF-int-rep}, resulting in
\begin{align} \label{eq:CF-FT}
\begin{split}
    \Phi^{(d)}_{\bs{N}}(\boldsymbol{\omega};u,\lambda)
    =&
    \frac{\Gamma\left(\frac{d+2}{2}\right)}{\pi^{\frac{d+2}{2}}} \lambda e^{\lambda\vert u \vert}    
     \\
     &
     \cdot 
     \int_{\mathbb{R}^{d+1}} 
     \big(
     \norm{\mathbf{n}}^2+\lambda^2
     \big)
     ^{-\frac{d+2}{2}}
    e^{
    i\mathbf{w}^\intercal \mathbf{n}
    }
    ~\mathrm{d}\mathbf{n},
\end{split}    
\end{align}
where
\color{black}
$\mathbf{n}=\left[\bs{n}^\tsp,\tilde{n}\right]^\tsp$ and $\mathbf{w}=\left[\boldsymbol{\omega}^\tsp,\tilde{\omega}\right]^\tsp$ are vectors in $\mathbb{R}^{d+1}$. 
\color{black}
We recognize that the integral in \eqref{eq:CF-FT} is proportional to the CF of a $(d+1)$-variate Cauchy distribution, with the last frequency variable $\tilde{\omega}$ fixed at $\vert u \vert$. The PDF and CF of a $(d+1)$-variate Cauchy vector $\mathbf{X}$ with location $\bs{\mu}=\bs{0}$ and scale $\bs{\Sigma}=\lambda^2 \mathbb{I}_{d+1}$  can be expressed as: \begin{align}
    f_\mathbf{X}(\mathbf{x})
    =
    \frac
    {\Gamma\left(\frac{d+2}{2}\right)}
    { \pi^{\frac{d+2}{2}} }
    \frac{\lambda}{\left( \|\mathbf{x}\|^2 + \lambda^2 \right)^{\frac{d+2}{2}} }
    ,
\end{align}
and
$
    \mathbb{E}\left[
    e^{i\mathbf{w}^\intercal\mathbf{X}}
    \right]
    =
    \exp
    (-\lambda\|\mathbf{w}\|),
$
as shown in \cite{lee2014clarification,kotz2004multivariate}.
Hence, using the FT pair relationship between the PDF and CF of $\mathbf{X}$, we can evaluate \eqref{eq:CF-FT} and get
\begin{align}
\begin{split}
    \Phi^{(d)}_{\bs{N}}(\boldsymbol{\omega})
    &=
    e^{\lambda|u|} \exp(-\lambda\|\mathbf{w}\|)
    \\&=
    \exp 
    \bigg(
    -\lambda 
    \Big( 
    \sqrt{ \norm{\boldsymbol{\omega}}^2 + \vert u\vert^2 } - \vert u \vert 
    \Big)
    \bigg)
    ,
\end{split}
\end{align}
where we made use of the substitution $\tilde{\omega}=|u|$.


\section{
\color{black}
Closed-form Calculation of the Differential Entropy for $d=2$ VDFAP Distribution
\color{black}
} \label{sec:appen-DE}

We directly calculate the differential entropy as the integral in the definition. After converting to polar coordinates (suggested by radial symmetry) and making a substitution, the integral reduces to the form of
the following integral formula\textcolor{black}{\cite[Section 8.432]{gradshteyn2014table}}: for $a>0$,
\begin{align}
\begin{split}
    &
    \int_a^\infty 
    \frac{K_{3/2}(\rho)}{\rho^{1/2}}
    \log
    \left(
    \frac{K_{3/2}(\rho)}{\rho^{3/2}}
    \right)
    \mathrm{d}\rho
    \\
    =~
    &
    \sqrt{\tfrac{\pi}{2}}
    \Big(
    e 
    \cdot 
    \text{Ei}(-1-a)
    -
    3
    \cdot
    \text{Ei}(-a)
    -
    e^{-a}
    \\
    &
    -
    (2a)^{-1} e^{-a}
    \left(
    6 \log a - 2 \log (1+a) + 6 + \log \tfrac{2}{\pi}
    \right)
    \Big)
    .
\label{eq:3DFAP-int-form}
\end{split}
\end{align}
Using this formula, we are able to write the differential entropy in a simplified closed-form as \eqref{eq:diff-ent-3D-VDFAP}.

\section{
\color{black}
Strictly Monotonically Increasing Property of the Ancillary Function $h_0(\cdot)$
\color{black}} 
\label{sec:appen-SI}

Observe that
$
h_0(s) = 2 \log(s) - \log(1+s) - g(s)
$,
where we introduced a function $g:\R^+ \to \R$ defined by
\begin{align}
    g(s)
    =
    s e^{s+1} 
    \text{Ei}(-(s+1))
    - 
    3 
    s e^s 
    \text{Ei}(-s),
    \text{  for  } s > 0
    .
\label{eq:def-g}
\end{align}
Taking the derivative of \eqref{eq:def-g} and using the definition of $\text{Ei}(\cdot)$ given in \eqref{eq:def-exp-int}, we obtain the following formula by direct differentiation. That is, 
\color{black}
\begin{align}
g'(s) = \dv{s} g(s)
=
\frac{s+1}{s} g(s) + \frac{s}{s+1} - 3
.
\end{align}
\color{black}
As a consequence, the derivative of $h_0(\cdot)$ can be expressed as
\begin{align}
    h_0'(s) = \dv{s} h_0(s)
    =
    \frac{s+1}{s} (2-g(s))
    .
\end{align}
\color{black}
To show that $h_0'(s)>0$ for any $s>0$, and hence $h_0(\cdot)$ is strictly monotonically increasing, it suffices to show that $g(s)<2$ for any $s>0$. Using the inequalities \cite[p.201]{luke1969special}:
\begin{align}
    \frac{s}{s+1} < -s e^s \text{Ei}(-s) < \frac{s+1}{s+2}
    \quad(\forall s > 0)
    ,
\end{align}
we can deduce that
\begin{align}
\begin{split}
    g(s)
    &<
    3 \cdot \frac{s+1}{s+2} - \frac{s}{s+1} \cdot \frac{s+1}{(s+1)+1}
    =
    \frac{2s+3}{s+2}
    <
    2,
\end{split}
\end{align}
which holds for any $s>0$.
Therefore, we have established the previously mentioned sufficient condition for $h_0(\cdot)$ to be strictly monotonically increasing.


\section{Calculating FAP Density in Specific Scenarios}\label{sec:case}

This section demonstrates the application of formula \eqref{eq:main} and the methodology presented in Section~\ref{section:generator} to compute the FAP density in some practical scenarios that are relevant to MC. 

\subsection{2D Linear Receiver and 3D Planar Receiver}
In this subsection, we unify the FAP density calculation for 2D linear receivers and 3D planar receivers by employing the innovative approach introduced in Section~\ref{section:generator}.

To calculate the $d$-dimensional FAP density, we consider an It\^{o} diffusion $\mathbf{X}_t\in\R^{\text{D}}$ with a semigroup generator
$
A := \sum_{i=1}^{\text{D}} v_i \dfrac{\partial}{\partial x_i} + \dfrac{\sigma^2}{2} \sum_{i=1}^{\text{D}} \dfrac{\partial^2}{\partial x_i^2}
$
(recall $d:=\text{D}-1$).
We denote the Laplacian operator\textcolor{black}{\cite[Section~8.5.1]{polyanin2001handbook}} in $\text{D}$ dimensions by $\Delta_\text{D}$ and consider the following drift-diffusion BVP in Cartesian coordinates:
\begin{equation}
\left\{\begin{array}{ll}
A\phi=\sum_{i=1}^{\text{D}} v_{i} \frac{\partial \phi}{\partial x_{i}}+\frac{\sigma^2}{2} \Delta_\text{D} \phi=0 & \text { in } \Omega \\
\phi=g & \text { on } \partial \Omega
\end{array}\right.
\label{BVP-DD},
\end{equation}
where $\phi$ is a (dummy) unknown function, $\Omega=\mathbb{R}^{\text{D}} \cap\left\{x_{\text{D}}>0\right\}$ is the domain of \eqref{BVP-DD}, and $\partial\Omega=\mathbb{R}^{\text{D}} \cap\left\{x_{\text{D}}=0\right\}$ is the boundary of the domain. Here, we use the notation $x_j$ to indicate the $j$-th component of the position vector $\mathbf{x}=[x_1,\ldots,x_\text{D}]^\tsp$ in an $\text{D}$-dimensional space. We temporarily set $\sigma^2=1$, and then treat the general case later.

To simplify the problem further, we introduce a change of variable
$
\psi(\mathbf{x})=\gamma(\mathbf{x}) \phi(\mathbf{x}).
$
The \emph{drift factor} $\gamma(\cdot)$ is defined as
$
\gamma(\mathbf{x}):=\exp {\mathbf{v}^\tsp \mathbf{x}},
$
which depends only on the drift components.
After changing the variables, we obtain a new BVP with the same $\Omega$ as 
\begin{equation}
\left\{\begin{array}{ll}
\tilde{A}(\psi)=\Delta_\text{D} \psi-s^{2} \psi=0 & \text { in } \Omega \\
\psi=\tilde{g} & \text { on } \partial \Omega
\end{array}\right.
,
\label{BVP-DD2}
\end{equation}
where
$
s=\|\mathbf{v}\|
$ denotes the magnitude of the drift vector.
This new operator $\tilde{A}$ is in the form of a Helmholtz operator (see\textcolor{black}{\cite[Section~7.3 and Section~8.3]{polyanin2001handbook}}), so we call this new BVP \eqref{BVP-DD2} the Helmholtz BVP.

According to \cite{polyanin2001handbook}, for both $\text{D}=2$ and $\text{D}=3$, the solution to the Helmholtz BVP \eqref{BVP-DD2} can be compactly written as
\begin{equation}
\psi(\mathbf{x})=\int_{\mathbf{y}\in \partial \Omega}\Big|\frac{\partial \tilde{G}(\mathbf{x}, \mathbf{y})}{\partial \mathbf{n}_\mathbf{y}}\Big| \tilde{g}(\mathbf{y}) ~\mathrm{d} \mathbf{y},
\label{eq:new-repreDD}
\end{equation}
where the notation $\big|\frac{\partial \tilde{G}(\mathbf{x}, \mathbf{y})}{\partial \mathbf{n}_\mathbf{y}}\big|$ denotes the absolute value of the directional derivative of $\tilde{G}$ with respect to the outward normal on the boundary $\partial \Omega$. The Green's function\footnote{We use $\tilde{G}$ to emphasize that this is the Green's function corresponding to the Helmholtz BVP, not to the original BVP.} $\tilde{G}$ corresponding to Helmholtz BVP \eqref{BVP-DD2} is known and can be separately written down for $\text{D}=2$ and $\text{D}=3$.

In order to obtain a representation formula for the solution $\phi$ of the original BVP \eqref{BVP-DD}, we substitute $\psi(\mathbf{x})=\gamma(\mathbf{x}) \phi(\mathbf{x})$ back into \eqref{eq:new-repreDD}, yielding
\begin{equation}
\begin{aligned}
\phi(\mathbf{x}) &= \frac{\psi(\mathbf{x})}{\gamma(\mathbf{x})} =\exp \{-\mathbf{v}^\tsp \mathbf{x}\} \int_{\partial \Omega}
\Big|\frac{\partial \tilde{G}(\mathbf{x}, \mathbf{y})}{\partial \mathbf{n}_{\mathbf{y}}}\Big| \tilde{g}(\mathbf{y}) ~\mathrm{d} \mathbf{y} \\
&\overset{(a)}{=}\int_{\partial \Omega} \exp \left\{\mathbf{v}^\tsp \mathbf{y}-\mathbf{v}^\tsp \mathbf{x}\right\}
\Big|\frac{\partial \tilde{G}(\mathbf{x}, \mathbf{y})}{\partial \mathbf{n}_{\mathbf{y}}}\Big|
g(\mathbf{y}) ~\mathrm{d} \mathbf{y} \\
&\overset{(b)}{=}\int_{\partial \Omega} \mathcal{K}_{\mathbf{v}}(\mathbf{x}, \mathbf{y}) g(\mathbf{y}) ~\mathrm{d} \mathbf{y}
\end{aligned}.
\label{e:calcul-01aD}
\end{equation}
In $(a)$, we have used a relation between the boundary data before and after the change of variable: 
$
\tilde{g}(\mathbf{y})=e^{\mathbf{v}^{\mathsf{T}}\mathbf{y}}g(\mathbf{y}).
$
In $(b)$, we use the notation $\mathcal{K}_{\mathbf{v}}(\mathbf{x}, \mathbf{y})$ to represent the overall ``integral kernel'' \textcolor{black}{(see \cite[Section~6.1]{kreyszig2011advanced} for this term).} Note that this integral kernel can be also expressed by $\big|\frac{\partial G(\mathbf{x}, \mathbf{y})}{\partial \mathbf{n}_\mathbf{y}}\big|$, where $G$ corresponds to the Green's function of original BVP \eqref{BVP-DD}. From \eqref{e:calcul-01aD}, we see that
\begin{align}
    \Big|\frac{\partial G(\mathbf{x}, \mathbf{y})}{\partial \mathbf{n}_\mathbf{y}}\Big|
    =
    \mathcal{K}_{\mathbf{v}}(\mathbf{x}, \mathbf{y})
    =
    \exp \left\{\mathbf{v}^\tsp \left( \mathbf{y}-\mathbf{x}\right) \right\}
\Big|\frac{\partial \tilde{G}(\mathbf{x}, \mathbf{y})}{\partial \mathbf{n}_{\mathbf{y}}}\Big|
.
\label{eq:FAP-density}
\end{align}

In the following, we separately write down, for $\text{D}=2$ and $\text{D}=3$, the Green's function $\tilde{G}$ corresponding to Helmholtz BVP \eqref{BVP-DD2} from \cite{polyanin2001handbook} and explicitly carry out the calculation in \eqref{eq:FAP-density}.

\subsubsection{2D Linear Receiver}\label{subsec:2D-line}

\color{black}
\color{black}

\color{black}
According to\textcolor{black}{\cite[Section~7.3]{polyanin2001handbook}},
when $\text{D}=2$,
the Green's function
corresponding to Helmholtz BVP \eqref{BVP-DD2} has the form
\begin{equation}\label{gf:2d}
\tilde{G}(\mathbf{x},\mathbf{y})
=\tilde{G}(x_1, x_2, \xi, \eta)=
\frac{1}{2\pi}[K_0(s\rho_1)-K_0(s\rho_2)]
,
\end{equation} 
where $\mathbf{x}=[x_1,x_2]^\tsp$ and $\mathbf{y}=[\xi,\eta]^\tsp$ are general positions in 2D, and we have defined
$
\rho_1:=\sqrt{(x_1-\xi)^{2}+(x_2-\eta)^{2}}
$
and 
$
\rho_2:=\sqrt{(x_2-\xi)^{2}+(x_2+\eta)^{2}}.
$
\color{black}
\color{black}
From \eqref{gf:2d}, we can directly calculate
\begin{align}
\begin{split}
\Big|\frac{\partial \tilde{G}(\mathbf{x}, \mathbf{y})}{\partial \mathbf{n}_{\mathbf{y}}}
\Big| \Bigg\vert_{\partial \Omega}
&=
\left[\dfrac{\partial \tilde{G}}{\partial \rho_1}\dfrac{\partial \rho_1}{\partial \eta}+\dfrac{\partial \tilde{G}}{\partial \rho_2}\dfrac{\partial \rho_2}{\partial \eta}\right]_{\eta=0}\\
&=
\dfrac{s\lambda}{\pi} \dfrac{K_1\left(s \sqrt{(x_1-\xi)^2+\lambda^2}\right)}{\sqrt{(x_1-\xi)^2+\lambda^2}},
\end{split}
\label{eq:tilde-G-deriv-2D}
\end{align}
where we have specified the emission point $\mathbf{x}=[x_1,\lambda]^\tsp$ as well as the arrival point $\mathbf{y}=[\xi,0]^\tsp$  hereafter for the 2D linear receiver. Notice that $\mathbf{y}$ now lies on the boundary $\partial \Omega$. The parameter $\lambda>0$ is the distance from the emission point to the linear receiver, as illustrated in Fig.~\ref{fig:2Da}.
Then according to \eqref{eq:FAP-density} and $s=\|\mathbf{v}\|$, we obtain
\begin{align}
\begin{split}
    \Big|\frac{\partial G(\mathbf{x}, \mathbf{y})}{\partial \mathbf{n}_\mathbf{y}}\Big |\Bigg\vert_{\partial \Omega}
    &=
    \mathcal{K}_{\mathbf{v}}(\mathbf{x}, \mathbf{y})
    \Big\vert_{\partial \Omega}\\
    &=
    \exp \left\{\mathbf{v}^\tsp \left( \mathbf{y}-\mathbf{x}\right) \right\}
\dfrac{\|\mathbf{v}\|\lambda}{\pi} \dfrac{K_1\left(\|\mathbf{v}\| \|\mathbf{y}-\mathbf{x}\|\right)}{\|\mathbf{y}-\mathbf{x}\|},
\end{split}
\end{align}
where 
we have substituted $\|\mathbf{y}-\mathbf{x}\|=\sqrt{(x_1-\xi)^2+\lambda^2}$ into \eqref{eq:tilde-G-deriv-2D}.
This $\mathcal{K}_{\mathbf{v}}(\mathbf{x}, \mathbf{y})\Big\vert_{\partial \Omega}$ is just the desired FAP density according to relation \eqref{eq:main} presented in Section~\ref{section:generator}.
We will denote this integral kernel by $f_{\mathbf{Y} \mid \mathbf{X}}(\mathbf{y}|\mathbf{x})$ hereafter.

For the general case $\sigma^2\neq 1$, we only need to replace $v_i$ with $\frac{v_i}{\sigma^2}=u_i$ (recall $\mathbf{u}:=\frac{\mathbf{v}}{\sigma^2}$), yielding the desired FAP density that allows arbitrary drift directions:
\begin{align}
\begin{split}
f_{\mathbf{Y} \mid \mathbf{X}}(\mathbf{y}|\mathbf{x})
=
\dfrac{\|\mathbf{u}\|\lambda}{\pi}
\exp \left\{\mathbf{u}^\tsp \left( \mathbf{y}-\mathbf{x}\right) \right\}
 \dfrac{K_1\left(\|\mathbf{u}\| \|\mathbf{y}-\mathbf{x}\|\right)}{\|\mathbf{y}-\mathbf{x}\|}
 .
\end{split}
\label{2Dresult}
\end{align}
Equation \eqref{2Dresult} is a new formula in the MC field.
Setting $v_1=0$ in \eqref{2Dresult} recovers the restrictive version in \cite{pandey2018molecular}:
\begin{align}
\begin{split}
&f_{\mathbf{Y} \mid \mathbf{X}}(\mathbf{y}|\mathbf{x})
\\
=~&\dfrac{|v_2|\lambda}{2\pi D}
\exp\left\{\dfrac{-v_2 \lambda}{2D}\right\} \dfrac{K_1\left(\dfrac{|v_2|}{2D}\sqrt{(x_1-\xi)^2+\lambda^2}\right)}{\sqrt{(x_1-\xi)^2+\lambda^2}}
,
\end{split}
\label{eq:2DFAP-corrected}
\end{align}
where we have used the relation $\sigma^2 = 2D$.\footnote{
In the literature, the diffusion coefficient $D$ may have different definitions, which can vary by a constant factor. However, the unit of $D$ consistently corresponds to $\sigma^2$.
}
To conclude, our formula \eqref{2Dresult} improves the formula \cite[eq.~(19)]{pandey2018molecular} by allowing arbitrary drift directions.
\color{black}

\color{black}
\subsubsection{3D Planar Receiver}
\color{black} 
According to\textcolor{black}{\cite[Section~8.3]{polyanin2001handbook}}, when $\text{D}=3$, the Green's function corresponding to Helmholtz BVP \eqref{BVP-DD2} has the form:
\begin{align}
\begin{split}
\tilde{G}(\mathbf{x},\mathbf{y}) &= \tilde{G}(x_1, x_2, x_3, \xi, \eta, \zeta)\\
&= \dfrac{\exp \left(-s \mathcal{R}_{1}\right)}{4 \pi \mathcal{R}_{1}}-\dfrac{\exp \left(-s \mathcal{R}_{2}\right)}{4 \pi \mathcal{R}_{2}}, 
\label{e:handbook3}
\end{split}
\end{align}
where $\mathbf{x}=[x_1,x_2,x_3]^\tsp$ and $\mathbf{y}=[\xi,\eta,\zeta]^\tsp$ are general positions in 3D, and we have defined 
$\mathcal{R}_1:=\sqrt{(x_1-\xi)^{2}+(x_2-\eta)^{2}+(x_3-\zeta)^{2}}$ and $\mathcal{R}_2:=\sqrt{(x_1-\xi)^{2}+(x_2-\eta)^{2}+(x_3+\zeta)^{2}}$.
From \eqref{e:handbook3}, 
we can directly differentiate to get
\begin{align}
\begin{split}
\Big|\frac{\partial \tilde{G}(\mathbf{x}, \mathbf{y})}{\partial \mathbf{n}_{\mathbf{y}}}\Big|
\Bigg\vert_{\partial \Omega}
&=
\left[\frac{\partial}{\partial \zeta} \tilde{G}(x_1,x_2,x_2,\xi,\eta,\zeta)\right]_{\zeta=0}\\
&=
\frac{\lambda}{2 \pi} \exp \{-s\|\mathbf{y}-\mathbf{x}\|\}
\left[\frac{1+s\|\mathbf{y}-\mathbf{x}\|}{\|\mathbf{y}-\mathbf{x}\|^{3}}\right]
\label{e:calcul-00}
\end{split}
\end{align}
where we have specified the emission point $\mathbf{x}=[x_1,x_2,\lambda]^\tsp$ and the arrival point $\mathbf{y}=[\xi,\eta,0]^\tsp$ hereafter for the 3D planar receiver. Notice that $\mathbf{y}$ now lies on the boundary $\partial \Omega$.
The parameter $\lambda>0$ is the distance from the emission point to the planar receiver, as illustrated in Fig.~\ref{fig:3Da}. 
Then according to \eqref{eq:FAP-density} and $s=\|\mathbf{v}\|$, we obtain
\begin{align}
\begin{split}
    &\Big|\frac{\partial G(\mathbf{x}, \mathbf{y})}{\partial \mathbf{n}_\mathbf{y}}\Big |\Bigg\vert_{\partial \Omega}
    =
    \mathcal{K}_{\mathbf{v}}(\mathbf{x}, \mathbf{y})
    \Big\vert_{\partial \Omega}
    \\
    =~&
    \exp \left\{\mathbf{v}^\tsp \left( \mathbf{y}-\mathbf{x}\right) \right\} \\
    & \frac{\lambda}{2 \pi} \exp \{-\|\mathbf{v}\|\|\mathbf{y}-\mathbf{x}\|\}
\left[\frac{1+\|\mathbf{v}\|\|\mathbf{y}-\mathbf{x}\|}{\|\mathbf{y}-\mathbf{x}\|^{3}}\right].
\end{split} 
\end{align}
This integral kernel $\mathcal{K}_{\mathbf{v}}(\mathbf{x}, \mathbf{y})
    \Big\vert_{\partial \Omega}$ can be interpreted as the probability density function of the FAP (see \eqref{eq:main}, Section~\ref{section:generator}), so we denote it by $f_{\mathbf{Y} \mid \mathbf{X}}(\mathbf{y}|\mathbf{x})$ hereafter.

Lastly, to treat the general case $\sigma^2\neq 1$, we can simply replace $v_i$ with $\frac{v_i}{\sigma^2}=u_i$ (recall $\mathbf{u}:=\frac{\mathbf{v}}{\sigma^2}$). This yields:
\begin{align}
\begin{split}
f_{\mathbf{Y} \mid \mathbf{X}}(\mathbf{y}|\mathbf{x})
=~&
\frac{\lambda}{2 \pi} 
\exp\left\{\mathbf{u}^\tsp \left(\mathbf{y}-\mathbf{x}\right)\right\}\\
&\exp \left\{-\|\mathbf{u}\|\norm{\mathbf{y}-\mathbf{x}}\right\}\left[\frac{1+\|\mathbf{u}\|\norm{\mathbf{y}-\mathbf{x}}}{\norm{\mathbf{y}-\mathbf{x}}^{3}}\right].
\label{e:2Dresult}
\end{split}
\end{align}
Equation~\eqref{e:2Dresult} is the desired 3D FAP density function, allowing arbitrary drift directions.
\color{black}
\subsection{3D Spherical Receiver}

\color{black}
In this subsection, we tackle the 3D spherical receiver case with the assumption of zero drift velocity, the same setting as in \cite{akdeniz2018molecular}. 
Note that the result we are going to show is already known in the literature, so we present a very brief version of how to obtain the angular density using our proposed methodology. The main goal of this subsection is to demonstrate the capability of our method to handle receivers with various shapes.
\color{black}

\color{green}
\color{black}

We consider an It\^{o} diffusion $\mathbf{X}_t\in\R^3$ with zero drift, i.e., all $v_i=0$. The corresponding generator $A$ is simply the 3D Laplacian operator. 
It is well known that the 3D Laplace equation in the spherical coordinate\textcolor{black}{
\cite[Section~12.11]{kreyszig2011advanced}} can be expressed as
\begin{align}
\begin{split}
\frac{1}{r^{2}} \frac{\partial}{\partial r}\left(r^{2} \frac{\partial \phi}{\partial r}\right)
&+\frac{1}{r^{2} \sin \theta} \frac{\partial}{\partial \theta}\left(\sin \theta \frac{\partial \phi}{\partial \theta}\right)\\
&+\frac{1}{r^{2} \sin ^{2} \theta} \frac{\partial^{2} \phi}{\partial \varphi^{2}}=0
\end{split}
\end{align}
where $r=\sqrt{x^2+y^2+z^2}$. Assuming a prescribed boundary data $g(\theta,\varphi)$ at the surface ($r=R$) of the spherical receiver with radius $R>0$, i.e.,
$
    \phi=g(\theta,\varphi) \text{\ at\ } r=R,
$
the solution of the outer problem for $r\geq R$ can be obtained as \cite[Section~8.1.3]{polyanin2001handbook}:
\begin{align}
\begin{split}
&\phi(r, \theta, \varphi)\\
=&\ \frac{R}{4 \pi} \int_{0}^{2 \pi} \int_{0}^{\pi} g\left(\theta_{0}, \varphi_{0}\right)\cdot\\ &\frac{r^{2}-R^{2}}{\left(r^{2}-2 rR \cdot \kappa(\theta,\varphi;\theta_0,\varphi_0)+R^{2}\right)^{3 / 2}} \sin \theta_{0} \diff \theta_{0} \diff \varphi_{0},
\end{split}
\label{eq:43}
\end{align}
where
\begin{align}
\kappa(\theta,\varphi;\theta_0,\varphi_0) = \cos\theta \cos\theta_0 + \sin\theta \sin\theta_0 \cos(\varphi-\varphi_0).
\end{align}
\color{black}


\color{black}
A comparison between formula \eqref{eq:43} and equation \eqref{general-green-2} allows us to determine
$\big| \frac{\partial G}{\partial \mathbf{n}_\mathbf{y}}(\mathbf{x},\mathbf{y}) \big|$, which is exactly the FAP density $f_{\mathbf{Y}|\mathbf{X}}(\mathbf{y}|\mathbf{x})$ due to our proposed formula \eqref{eq:main}. That is, we have already obtained the (marginal) \emph{angular density} of the hitting position of molecules, which was expressed in \cite[eq.~(3)-(4)]{akdeniz2018molecular} and \cite[eq.~(6.3.3a)]{redner2001guide}. This demonstrates the versatility of our FAP calculating methodology in handling various receiver shapes, including spherical receivers.
\color{black}




\bibliographystyle{IEEEtranTCOM}
\balance
\bibliography{main}
\end{document}